\documentclass[11pt]{article}
\begin{document}
\setcounter{page}{1}
\def\theequation{\arabic{section}.\arabic{equation}}
\def\theequation{\thesection.\arabic{equation}}
\setcounter{section}{0}

\title{On the path integral representation for Wilson loops and the
non--Abelian Stokes theorem\thanks{The work supported in part by
Jubil\"aumsfonds of the Austrian National Bank Project No.7232}}

\author{M. Faber\thanks{E--mail: faber@kph.tuwien.ac.at, Tel.:
+43--1--58801--14261, Fax: +43--1--58801--14299} ,
A. N. Ivanov\thanks{E--mail: ivanov@kph.tuwien.ac.at, Tel.:
+43--1--58801--14261, Fax: +43--1--58801--14299}~$^{\S}$ , N. I.
Troitskaya\thanks{Permanent Address: State
Technical University, Department of Nuclear Physics, 195251
St. Petersburg, Russian Federation} , M.  Zach\thanks{E--mail:
zach@kph.tuwien.ac.at, Tel.: +43--1--58801--14266, Fax:
+43--1--58801--14299}}

\date{\today}

\maketitle

\begin{center}
{\it Institut f\"ur Kernphysik, Technische Universit\"at Wien, \\ 
Wiedner Hauptstr. 8--10, A--1040 Vienna, Austria}
\end{center}

\begin{abstract}
We discuss the derivation of the path integral representation over
gauge degrees of freedom for Wilson loops in $SU(N)$ gauge theory and
4--dimensional Euclidean space--time by using well--known properties
of group characters. A discretized form of the path integral is
naturally provided by the properties of group characters and does not
need any artificial regularization. We show that the path integral
over gauge degrees of freedom for Wilson loops derived by Diakonov and
Petrov (Phys. Lett. B224 (1989) 131) by using a special regularization
is erroneous and predicts zero for the Wilson loop. This property is
obtained by direct evaluation of path integrals for Wilson loops
defined for pure $SU(2)$  gauge fields and $Z(2)$ center vortices with
spatial azimuthal symmetry. Further we show that both derivations
given by Diakonov and Petrov for their regularized path integral, if
done correctly, predict also zero for Wilson loops. Therefore, the
application of their path integral representation of Wilson loops
cannot give {\it a new way to check confinement in lattice} as has
been declared by Diakonov and Petrov (Phys. Lett. B242 (1990)
425). From the path integral representation which we consider we
conclude that no new non--Abelian Stokes theorem can exist for Wilson
loops except the old--fashioned one derived by means of the
path-ordering procedure.
\end{abstract}

\begin{center}
\end{center}

\newpage

\section{Introduction}
\setcounter{equation}{0}

\hspace{0.2in} The hypothesis of quark confinement, bridging the
hypothesis of the existence of quarks and the failure of the detection
of quarks as isolated objects, is a challenge for QCD. As a criterion
of colour confinement in QCD, Wilson [1] suggested to consider the
average value of an operator
\begin{eqnarray}\label{label1.1}
W(C) = \frac{1}{N}\,{\rm tr}\,{\cal P}_C\,e^{\textstyle i\,g\,\oint_C d
x_{\mu}\,A_{\mu}(x)} = \frac{1}{N}\,{\rm tr}\, U(C_{x x}),
\end{eqnarray}
defined on an closed loop $C$, where $A_{\mu}(x) = t^a\,A^a_{\mu}(x)$
is a gauge field, $t^a$ ($a = 1,\ldots, N^2-1$) are the generators of
the $SU(N)$ gauge group in fundamental representation normalized by
the condition ${\rm tr}\,(t^a t^b) = \delta^{ab}/2$, $g$ is the gauge
coupling constant and ${\cal P}_C$ is the
operator ordering colour matrices along the path $C$. The trace in Eq.(\ref{label1.1}) is computed over colour
indices. The operator
\begin{eqnarray}\label{label1.2}
U(C_{y x}) = {\cal P}_{C_{y x}}e^{\textstyle i\,g\,\int_{C_{y x}} d
z_{\mu}\,A_{\mu}(z)},
\end{eqnarray}
makes a parallel transport along the path $C_{y x}$ from $x$ to $y$. For
Wilson loops the contour $C$ defines a closed path $C_{x
x}$.  For determinations of the parallel transport
operator $U(C_{y x})$ the action of the path--ordering operator ${\cal
P}_{C_{y x}}$ is defined by the following limiting procedure [2]
\begin{eqnarray}\label{label1.3}
\hspace{-0.5in}&&U(C_{y x})={\cal P}_{C_{xy}}e^{\textstyle
i\,g\,\int_{C_{y x}} d z_{\mu}\,A_{\mu}(z)}= \lim_{n \to
\infty}\prod^{n}_{k = 1}U(C_{x_k x_{k-1}}) =\nonumber\\
\hspace{-0.5in}&&= \lim_{n \to \infty}U(C_{y x_{n -
1}})\,\ldots\,U(C_{x_2 x_1})\,U(C_{x_1 x})=  \lim_{n \to \infty} \prod^{n}_{k =
1}e^{\textstyle i\,g\,(x_k-x_{k-1})\cdot A(x_{k-1})},
\end{eqnarray}
where $C_{x_k x_{k-1}}$ is an infinitesimal segment of the path $C_{y
x}$ with $x_{0} = x$ and
$x_{n} = y$. The parallel transport operator $U(C_{x_k x_{k-1}})$ for an infinitesimal segment $C_{x_k x_{k-1}}$ is defined by [2]:
\begin{eqnarray}\label{label1.4}
\hspace{-0.2in}U(C_{x_k x_{k-1}})= e^{\textstyle i\,g\,\int_{C_{x_k
x_{k-1}}} d z_{\mu}\,A_{\mu}(z)}=e^{\textstyle i\,g\,(x_k -
x_{k-1})\cdot A(x_{k-1})}.
\end{eqnarray}
In accordance with the definition of the path--ordering procedure
(\ref{label1.3}) the parallel transport operator $U(C_{y x})$ has the property
\begin{eqnarray}\label{label1.5}
U(C_{y x}) = U(C_{y x_1})\,U(C_{x_1 x}),
\end{eqnarray}
where $x_1$ belongs to the path $C_{y x}$. Under gauge transformations with a gauge function $\Omega(z)$,
\begin{eqnarray}\label{label1.6}
A_{\mu}(z) \to A^{\Omega}_{\mu}(z) = \Omega(z) A_{\mu}(z)
\Omega^{\dagger}(z) + \frac{1}{ig}\,\partial_{\mu}\Omega(z) 
\Omega^{\dagger}(z),
\end{eqnarray}
the operator $U(C_{y x})$ has a very simple transformation law
\begin{eqnarray}\label{label1.7}
U(C_{y x}) \to U^{\Omega}(C_{y x}) = \Omega(y)\,U(C_{y x})
\,\Omega^{\dagger}(x).
\end{eqnarray}
We would like to stress that this equation is valid even if the gauge
functions $\Omega(x)$ and $\Omega(y)$ differ significantly for
adjacent points $x$ and $y$.

As has been postulated by Wilson [1] the average value of the Wilson
loop $<W(C)>$ in the confinement regime should show area--law falloff
[1]
\begin{eqnarray}\label{label1.8}
<W(C)> \sim e^{\textstyle - \sigma\,{\cal A}},
\end{eqnarray}
where $\sigma$ and ${\cal A}$ are the string tension and the minimal
area of the loop, respectively. As usually the minimal area is a
rectangle of size $L\times T$.  In this case the exponent $\sigma
{\cal A}$ can be represented in the equivalent form $\sigma\,{\cal A}
= V(L)\,T$, where $V(L) = \sigma L$ is the interquark potential and
$L$ is the relative distance between quark and anti--quark.

The paper is organized as follows. In Sect.\,2 we discuss the path
integral representation for Wilson loops by using well--known
properties of group characters. The discretized form of this path
integral is naturally provided by properties of group characters and
does not need any artificial regularization. We derive a closed
expression for Wilson loops in irreducible representation $j$ of
 $SU(2)$. In Sect.\,3 we extend the path integral
representation to the gauge group $SU(N)$. As an example, we give an
explicit representation for Wilson loops in the fundamental
representation of $SU(3)$. In Sects.\,4 and 5 we evaluate the path
integral for Wilson loops, suggested in Ref.[3], for two specific
gauge field configurations (i) a pure gauge field in the fundamental
representation of $SU(2)$ and (ii) $Z(2)$ center vortices with spatial
azimuthal symmetry, respectively. We show that this path integral
representation fails to describe the original Wilson loop for both
cases. In Sect.\,6 we show that the regularized evolution operator in
Ref.[3] representing Wilson loops in the form of the path integral
over gauge degrees of freedom has been evaluated incorrectly by
Diakonov and Petrov. The correct value for the evolution operator is
zero. This result agrees with those obtained in Sects.\,4 and 5. In
Sect.\,7 we criticize the removal of the oscillating factor from the
evolution operator suggested in Ref.[3] via a shift of energy levels
of the axial--symmetric top. We show that such a removal is
prohibited. It leads to a change of symmetry of the starting
system from $SU(2)$ to $U(2)$. Keeping the oscillating factor one gets
a vanishing value of Wilson loops in agreement with our results in
Sects.\,4, 5 and 6. In the Appendix we evaluate the coefficients of the expansion used for the path integrals in Sects.\,4 and 5.

\section{Path integral representation for Wilson loops}
\setcounter{equation}{0}

 Attempts to derive a path integral representation for Wilson loops
 (\ref{label1.1}), where the path ordering operator is replaced by a path integral, have been undertaken in Refs.[3--5]. The path integral representations have been derived for Wilson loops in terms of gauge degrees of freedom (bosonic variables) [3,4] and fermionic
 degrees of freedom (Grassmann variables) [5]. For the derivation of
 the quoted path integral representations for Wilson loop different mathematical machineries have been used. Below we discuss the derivation of the path integral representation for Wilson loops in terms of gauge degrees of freedom by using well--known properties of group characters. In this case a discretized form of path integrals is naturally
 provided by the properties of group characters and the completeness
 condition of gauge functions. It coincides with the standard
 discretization of Feynman path integrals [6] and does not
 need any artificial regularization.

We argue that the path integral representation for Wilson loops
 suggested by Diakonov and Petrov in Ref.[3] is erroneous. For the
 derivation of this path integral representation Diakonov and Petrov
 have used a special regularization drawing an analogy with an
 axial--symmetric top. The moments of inertia of this top are taken finally to zero.
 As we show below this path integral amounts to zero for Wilson
 loops defined for $SU(2)$. Therefore, it is not a surprise that the application of this
 erroneous path integral representation to the evaluation of the
 average value of Wilson loops has led to the conclusion that for
 large loops the area--law falloff is present for colour charges taken
 in any irreducible representation $r$ of $SU(N)$ [7].  This statement
 has not been supported by numerical simulations within lattice QCD
 [8]. As has been verified, e.g. in Ref.[8] for $SU(3)$,  in the confined phase and  at large distances, colour charges with non--zero $N$--ality have string tensions of the corresponding fundamental representation, whereas colour charges with zero $N$--ality are screened by gluons and cannot form a string at large distances. Hence, the results obtained in Ref.[6] cannot give {\it a new way to check confinement in lattice} as has been declared by Diakonov and Petrov.

For the derivation of Wilson loops in the form of a path integral
 over gauge degrees of freedom by using well--known properties of 
 group characters it is convenient to represent $W(C)$ in terms of 
characters of irreducible representations of $SU(N)$ [9--11]
\begin{eqnarray}\label{label2.1}
W_r(C) = \frac{1}{\displaystyle  d_r}\,\chi[U_r(C_{x x})],
\end{eqnarray}
where the matrix $U_r(C_{x x})$ realizes an irreducible and
$d_r$--dimensional matrix representation $r$ of
the group $SU(N)$ with the character $\chi[U_r(C_{x x})] = {\rm
tr}[U_r(C_{x x})]$.  

In order to introduce the path integral over gauge degrees of freedom we
suggest to use 
\begin{eqnarray}\label{label2.2}
\int D\Omega_r\chi[U_r\Omega^{\dagger}_r]\,\chi[\Omega_rV_r]  =
\frac{1}{\displaystyle d_r}\,\chi[U_rV_r],
\end{eqnarray}
where the matrices $U_r$ and $V_r$ belong to the irreducible
representation $r$, and $D\Omega_r$ is the Haar
measure normalized to unity  $\int D\Omega_r = 1$. The completeness 
condition for gauge functions $\Omega_r$ reads
\begin{eqnarray}\label{label2.3}
\int D\Omega_r{(\Omega^{\dagger}_r)}_{\displaystyle a_1 b_1}
{(\Omega_r)}_{\displaystyle a_2 b_2} = \frac{1}{\displaystyle
d_r}\,\delta_{\displaystyle a_1 b_2}\,\delta_{\displaystyle b_1 a_2}.
\end{eqnarray}
By using the completeness condition it is convenient to represent the
Wilson loop in the form of the integral
\begin{eqnarray}\label{label2.4}
W_r(C) = \frac{1}{\displaystyle d_r}\,\int
D\Omega_r(x)\chi[\Omega_r(x)U_r(C_{x
x})\Omega^{\dagger}_r(x)].
\end{eqnarray}
According to Eq.(\ref{label1.3}) and Eq.(\ref{label1.5}) the matrix
$U_r(C_{x x})$ can be decomposed in
\begin{eqnarray}\label{label2.5}
U_r(C_{x x}) =\lim_{n\to \infty} U_r(C_{x x_{n-1}})
U_r(C_{x_{n-1} x_{n-2}})\ldots U_r(C_{x_2
x_1})U_r(C_{x_1 x}).
\end{eqnarray}
Substituting Eq.(\ref{label2.5}) in Eq.(\ref{label2.4}) and applying
$(n-1)$--times Eq.(\ref{label2.2}) we end up with
\begin{eqnarray}\label{label2.6}
W_r(C)&=& \frac{1}{d^2_r}\,\lim_{n\to \infty}\int\ldots\int D
\Omega_r(x_1) \ldots \Omega_r(x_n)\,d_r\chi[\Omega_r(x_n)U_r(C_{x_n
x_{n-1}}) \Omega^{\dagger}_r(x_{n-1})]\nonumber\\ &&\ldots
d_r\chi[\Omega_r(x_1) U_r(C_{x_1 x_n}) \Omega^{\dagger}_r(x_n)].
\end{eqnarray}
Using relations $\Omega_r(x_k) U_r(C_{x_k x_{k-1}})
\Omega^{\dagger}_r(x_{k-1}) = U^{\Omega}_r(C_{x_k x_{k-1}})$ we get
\begin{eqnarray}\label{label2.7}
W_r(C)=\frac{1}{d^2_r}\,\lim_{n\to \infty}\int \ldots \int D
\Omega_r(x_1)\ldots D \Omega_r(x_n)\, d_r\chi[U^{\Omega}_r(C_{x_n
x_{n-1}})]\ldots d_r\chi[U^{\Omega}_r(C_{x_1 x_n})].
\end{eqnarray}
The integrations over $\Omega_r(x_k)\,$ ($k=1,\ldots,n$) are well
defined. These are  standard integrations on the compact Lie group
$SU(N)$.

We should emphasize that the integrations over $\Omega_r(x_k)\,$
($k=1,\ldots,n$) are not correlated and should be carried out
independently. 

Since Eq.(\ref{label2.3}) is the completeness condition for group elements,
the discretization of Wilson loops given by Eqs.(\ref{label2.6}) and
(\ref{label2.7}) reproduces the standard discretization of Feynman path integrals [6] where infinitesimal time steps can be described by a classical motion. Therefore, the discretized expression (\ref{label2.7}) can be represented formally by
\begin{eqnarray}\label{label2.8}
W_r(C) = \frac{1}{d^2_r}\int \prod_{x\in C} \left[ d_r\,D
 \Omega_r(x)\right] \, \chi[U^{\Omega}_r(C_{x x})].
\end{eqnarray}
Conversely the evaluation of this path integral corresponds to the
discretization given by Eqs.(\ref{label2.6}) and (\ref{label2.7}).
The measure of the integration over $\Omega_r(x)$ is well defined and
normalized to unity
\begin{eqnarray}\label{label2.9}
\int \prod_{x\in C} D \Omega_r(x) = \lim_{n\to \infty}
\int D \Omega_r(x_n)\int D \Omega_r(x_{n - 1})\ldots 
\int D \Omega_r(x_1) = 1.
\end{eqnarray}
Thus, for the determination of the path integral over gauge degrees of
freedom (\ref{label2.8}) we do not need to use any regularization,
since the discretization given by Eqs.(\ref{label2.6}) and
(\ref{label2.7}) are well defined.

We would like to emphasize that Eq.(\ref{label2.8}) is a continuum
analogy of the lattice version of the path integral over gauge degrees
of freedom for Wilson loops used in Eq.(2.13) of Ref.[11] for the evaluation of the average value of Wilson loops in connection with $Z(2)$ center vortices. 

Now let us to proceed to the evaluation of the characters
$\chi[U^{\Omega}_r(C_{x_k x_{k-1}})]$.  Due to the infinitesimality of
the segments $C_{x_k x_{k-1}}$ we can omit the path ordering operator
in the definition of $U^{\Omega}_r(C_{x_k x_{k-1}})$ [2]. This allows
us to evaluate the character $\chi[U^{\Omega}_r(C_{x_k x_{k-1}})]$
with $U^{\Omega}_r(C_{x_k x_{k-1}})$ taken in the form [2]
\begin{eqnarray}\label{label2.10}
 U^{\Omega}_r(C_{x_k x_{k-1}}) = \exp ig\int_{C_{x_k
 x_{k-1}}}dx_{\mu}A^{\Omega}_{\mu}(x).
\end{eqnarray}
Of course, the relation given by Eq.(\ref{label2.10}) is only defined
in the sense of a meanvalue over an infinitesimal segment $C_{x_k
x_{k-1}}$. Therefore, it can be regarded to some extent as a
smoothness condition.  Unlike the smoothness condition used by
Diakonov and Petrov [3] Eq.(\ref{label2.10}) does not corrupt the
Wilson loop represented by the path integral over the gauge degrees of
freedom.
 
The evaluation of the characters of $U^{\Omega}_r(C_{x_k x_{k-1}})$
given by Eq.(\ref{label2.10}) runs as follows. First let us consider
the simplest case, the $SU(2)$ gauge group, where we have 
$r = j = 0,1/2,1,\ldots$ and $d_j = 2j + 1$. The
character $\chi [U^{\Omega}_j(C_{x_k x_{k-1}})]$ is equal to [9,10,12]
\begin{eqnarray}\label{label2.11}
\chi [U^{\Omega}_j(C_{x_k x_{k-1}})] &=& \sum^{j}_{m_j =-j}
<m_j|U^{\Omega}_j(C_{x_k x_{k-1}})|m_j> = \nonumber\\
&=&\sum^{j}_{m_j =-j} e^{\textstyle 
i\,m_j\, \Phi[C_{x_k x_{k-1}}; A^{\Omega}]},
\end{eqnarray}
where $m_j$ is the magnetic colour quantum number, $|m_j>$ and
$m_j\,\Phi [C_{x_k x_{k-1}}; A^{\Omega}]$ are the eigenstates and
eigenvalues of the operator
\begin{eqnarray}\label{label2.12}
\hat{\Phi}[C_{x_k x_{k-1}}; A^{\Omega}] = g\int_{C_{x_k
x_{k-1}}}dx_{\mu}A^{\Omega}_{\mu}(x),
\end{eqnarray}
i.e. $\hat{\Phi}[C_{x_k x_{k-1}};
A^{\Omega}]\,|m_j>\,=\,m_j\,\Phi[C_{x_k x_{k-1}};
A^{\Omega}]\,|m_j(x_{k-1})>$. The standard procedure for the
evaluation of the eigenvalues gives $\Phi[C_{x_k x_{k-1}};
A^{\Omega}]$ in the form
\begin{eqnarray}\label{label2.13}
\Phi[C_{x_k x_{k-1}}; A^{\Omega}] = g \int_{C_{x_k
x_{k-1}}}\sqrt{\displaystyle g_{\mu\nu}[A^{\Omega}](x)
dx_{\mu}dx_{\nu}},
\end{eqnarray}
where the metric tensor can be given formally by the expression
\begin{eqnarray}\label{label2.14}
g_{\mu\nu}[A^{\Omega}](x) =
2\,{\rm tr} [A^{\Omega}_{\mu}A^{\Omega}_{\nu}](x).
\end{eqnarray}
In order to find an explicit expression for the metric tensor we
should fix a gauge. As an example let us take the Fock--Schwinger
gauge
\begin{eqnarray}\label{label2.15}
x_{\mu}A_{\mu}(x) = 0.
\end{eqnarray}
In this case the gauge field $A_{\mu}(x)$ can be expressed in terms of
the field strength tensor $G_{\mu\nu}(x)$ as follows
\begin{eqnarray}\label{label2.16}
A_{\mu}(x) = \int\limits^1_0 ds\, s\, x_{\alpha}\, G_{\alpha\mu}(x s).
\end{eqnarray}
This can be proven by using the obvious relation
\begin{eqnarray}\label{label2.17}
\hspace{-0.2in}x_{\alpha} G_{\alpha\mu}(x)&=& 
x_{\alpha} \partial_{\alpha}A_{\mu}(x)
- x_{\alpha}\partial_{\mu}A_{\alpha}(x) - i g
[x_{\alpha}A_{\alpha}(x),A_{\mu}(x)] =\nonumber\\
&=& A_{\mu}(x) +
x_{\alpha}\frac{\partial}{\partial x_{\alpha}}A_{\mu}(x),
\end{eqnarray}
valid for the Fock--Schwinger gauge $x_{\alpha}A_{\alpha}(x) = 0$.
 Replacing $x \to x s$ we can represent the r.h.s. of
 Eq.(\ref{label2.17}) as a total derivative with respect to $s$
\begin{eqnarray}\label{label2.18}
s x_{\alpha} G_{\alpha\mu}(x s) = A_{\mu}(x s) +
x_{\alpha}\frac{\partial}{\partial x_{\alpha}}A_{\mu}(x s) = \frac{d}{d
s}[s A_{\mu}(x s)].
\end{eqnarray}
Integrating out $s \in [0,1]$ we arrive at Eq.(\ref{label2.16}).

Using Eq.(\ref{label2.16}) we obtain the metric tensor
$g_{\mu\nu}[A^{\Omega}](x)$ in the form
\begin{eqnarray}\label{label2.19}
&& g_{\mu\nu}[A^{\Omega}](x) =  2 x_{\alpha} x_{\beta}
\int\limits^1_0
\int\limits^1_0 ds ds' s s' 
{\rm tr}[G^{\Omega}_{\alpha\mu}(x
s)G^{\Omega}_{\beta\nu}(x s'\,)]=\nonumber\\ 
&&= 2 x_{\alpha}
x_{\beta}\int\limits^1_0 \int\limits^1_0 ds ds' s s' 
{\rm tr}[\Omega(x
s)G_{\alpha\mu}(x s)\Omega^{\dagger}(x s)
\Omega(x s'\,)G_{\beta\nu}(x
s'\,)\Omega^{\dagger}(x s'\,)].
\end{eqnarray}
For the derivation of Eq.(\ref{label2.19}) we define the operator
$\Phi[C_{x_k x_{k-1}}; A^{\Omega}]$ of Eq.(\ref{label2.12}) following the
definition of the phase of the parallel transport operator $U(C_{x_k
x_{k-1}})$ given by Eq.(\ref{label1.4}) [2]
\begin{eqnarray}\label{label2.20}
\hat{\Phi}[C_{x_k x_{k-1}}; A^{\Omega}] &=& g\int_{C_{x_k
x_{k-1}}}dx_{\mu}A^{\Omega}_{\mu}(x)=(x_k - x_{k-1})_{\mu}
A^{\Omega}_{\mu}(x_{k-1})=\nonumber\\ &=&(x_k -
x_{k-1})_{\mu}\int\limits^1_0 ds s\,x^{\alpha}_{k-1}
\,G^{\Omega}_{\alpha\mu}(x_{k-1} s).
\end{eqnarray}
The parameter $s$ is to some extent an order parameter distinguishing
the gauge functions $\Omega(x_k)$ and $\Omega(x_{k-1})$ entering
the relation $\Omega(x_k)U(C_{x_k x_{k-1}})\Omega^{\dagger}(x_{k-1}) =
U^{\Omega}(C_{x_k x_{k-1}})$. 

Substituting Eq.(\ref{label2.11}) in Eq.(\ref{label2.7}) we arrive at
the expression for Wilson loops defined for $SU(2)$ 
\begin{eqnarray}\label{label2.21}
\hspace{-0.3in}&&W_j(C)
=\frac{1}{(2j+1)^2}\lim_{n\to\infty}\nonumber\\ &&\int D
\Omega_j(x_n)\,(2j+1)\sum^{j}_{m^{(n)}_j=-j}\!\!\!e^{\textstyle i\,
 g\, m^{(n)}_j
\int_{C_{x_1
x_n}}\sqrt{g_{\mu\nu}[A^{\Omega}](x)\,dx_{\mu}dx_{\nu}}}\nonumber\\
\hspace{-0.3in}&&\int D \Omega_j(x_{n-1})\,(2j+1)\sum^{j}_{m^{(n-1)}_j
= -j}\!\!\!  e^{\textstyle i\, g\, m^{(n-1)}_j \int_{C_{x_n
x_{n-1}}}\sqrt{g_{\mu\nu}[A^{\Omega}](x)\,dx_{\mu}dx_{\nu}}}
\nonumber\\ &&\vdots\nonumber\\
\hspace{-0.3in}&&\int D \Omega_j(x_1)\,(2j+1)\sum^{j}_{m^{(1)}_j =
-j}\!\!\!  e^{\textstyle i\, g\, m^{(1)}_j \int_{C_{x_2 x_1}} 
\sqrt{g_{\mu\nu}[A^{\Omega}](x)\,dx_{\mu}dx_{\nu}}}.
\end{eqnarray}
The magnetic quantum number $m^{(k)}_j\,(k=1, \ldots, n)$ belongs to
the infinitesimal segment $C_{x_{k+1} x_k}$, where $C_{x_{n+1} x_n} =
C_{x_1 x_n}$.

In compact form Eq.(\ref{label2.21}) can be written as a path
integral over gauge functions
\begin{eqnarray}\label{label2.22}
\hspace{-0.2in}W_j(C)= \frac{1}{(2j+1)^2}\int \prod_{x\in C} D \Omega_j(x) \, \sum_{\left\{ m_j(x) \right\} }\!\!\!(2j+1)\,e^{\textstyle i g
\oint_{C} m_j(x)\,\sqrt{g_{\mu\nu}[A^{\Omega}](x)\,dx_{\mu}dx_{\nu}}}.
\end{eqnarray}
The integrals along the infinitesimal segments $C_{x_k x_{k-1}}$ we
determine as [2]
\begin{eqnarray}\label{label2.23}
\int_{C_{x_k x_{k-1}}}\!\!\!
m_j(x)\,\sqrt{g_{\mu\nu}[A^{\Omega}](x)\,dx_{\mu}dx_{\nu}}&=&
m_j(x_{k-1})\, \sqrt{g_{\mu\nu}[A^{\Omega}](x_{k-1})\,\Delta
x_{\mu}\,\Delta x_{\nu}}=\nonumber\\
&=& m^{(k-1)}_j\sqrt{g_{\mu\nu}[A^{\Omega}](x_{k-1})\,\Delta
x_{\mu}\,\Delta x_{\nu}}.
\end{eqnarray}
where $\Delta x = x_k - x_{k-1}$.

Comparing the path integral (\ref{label2.22}) with that
suggested in Eq.(23) of Ref.[3] one finds rather strong
disagreement. First, this concerns the contribution of different
states $m_j$ of the representation $j$. In the case of the path
integral (\ref{label2.22}) there is a summation over all values of
the magnetic colour quantum number $m_j$, whereas the representation
of Ref.[3] contains only one term with $m_j = j$. Second, Ref.[3] claims that in the integrand of their path integral the
exponent should depend only on the gauge field projected onto the
third axis in colour space. However, this is only possible
if the gauge functions are slowly varying with $x$,
i.e. $\Omega(x_k)\Omega^{\dagger}(x_{k-1})\simeq 1$.  In this case the
parallel transport operator $U^{\Omega}(C_{x_k x_{k-1}})$ would read
[13]
\begin{eqnarray}\label{label2.24}
U^{\Omega}(C_{x_k x_{k-1}})= \exp{i\,g\,\int_{C_{x_k x_{k-1}}} d
x_{\mu}\,A^{\Omega}_{\mu}(x)} = 1 + i\,g\,(x_i - x_{i-1})\cdot
A^{\Omega}(x_{i-1}),
\end{eqnarray}
and the evaluation of the character $\chi[U^{\Omega}_j(C_{x_k
x_{k-1}}]$ would run as follows
\begin{eqnarray}\label{label2.25}
\hspace{-0.3in}&&<m_j|[U^{\Omega}_j(C_{x_k x_{k-1}})]|m_j> =
 1 +(t^a_j)_{m_j m_j}\,
i\,g\,(x_k - x_{k-1})\cdot [A^{\Omega}(x_{k-1})]^{(a)}=\nonumber\\
\hspace{-0.3in}&&= 1 + m_j\,i\,g\,(x_k - x_{k-1})\cdot
[A^{\Omega}(x_{k-1})]^{(3)} = e^{\textstyle 
i\,g\int_{C_{x_k x_{k-1}}}dx_{\mu}\,m_j(x)[A^{\Omega}_{\mu}(x)]^{(3)}}\!\!\!,
\end{eqnarray}
where we have used the matrix elements of the generators of
$SU(2)$, i.e. $(t^a_j)_{m_j m_j} =
m_j\,\delta^{a3}$. More generally the exponent on the
r.h.s. of Eq.(\ref{label2.25}) can be written as
\begin{eqnarray}\label{label2.26}
\int_{C_{x_k x_{k-1}}}dx_{\mu}\,m_j(x)\,[A^{\Omega}_{\mu}(x)]^{(3)} =
2\int_{C_{x_k x_{k-1}}}dx_{\mu}\,m_j(x)\,{\rm tr}[t^3_j A^{\Omega}_{\mu}(x)].
\end{eqnarray}
This gives the path integral representation for Wilson loops
defined for $SU(2)$ in the following form
\begin{eqnarray}\label{label2.27}
\hspace{-0.3in}&&W_j(C)=\frac{1}{(2j+1)^2}\int \prod_{x\in C} D \Omega_j(x) \, \sum_{\left\{ m_j(x) \right\} }\!\!\!(2j+1)\,
e^{\textstyle 2 i  g \oint_{C}dx_{\mu}\,m_j(x)\,{\rm tr}[t^3_j A^{\Omega}_{\mu}(x)]}.\nonumber\\
\hspace{-0.3in}&&
\end{eqnarray}
The exponent contains the gauge field projected onto the third axis in
colour space ${\rm tr}[t^3_j A^{\Omega}_{\mu}(x)]$.
 Nevertheless, Eq.(\ref{label2.27}) differs form Eq.(23)
of Ref.[3] by a summation over all values of the colour magnetic quantum
number $m_j$ of the given irreducible representation $j$.

The repeated application of Eq.(\ref{label2.2}) induces that the
integrations over the gauge function at $x_k$ are completely
independent of the integrations at $x_{k\pm 1}$. There is no mechanism
which leads to gauge functions smoothly varying with $x_k\,(k =
1,\ldots\,n)$. In this sense the situation is opposite to the quantum
mechanical path integral. In Quantum Mechanics the integration over
all paths is restricted by the kinetic term of the Lagrange function.
In the semiclassical limit $\hbar \to 0$ due to the kinetic term the
fluctuations of all trajectories are shrunk to zero around a
classical trajectory. However, in the case of the integration over
gauge functions for the path integral representation of the Wilson
loop, there is neither a suppression factor nor a semiclassical limit
like $\hbar \to 0$. The key point of the application of
Eq.(\ref{label2.2}) and, therefore, the path integral representation
for Wilson loops is that all integrations over $\Omega(x_k)\,(k =
1,\ldots,n)$ are completely independent and can differ substantially
even if the points, where the gauge functions $\Omega(x_k)$ and
$\Omega(x_{k-1})$ are defined, are infinitesimally close to each other.

For the derivation of Eq.(23) of Ref.[3] Diakonov and Petrov have used
at an intermediate step a regularization drawing an analogy with an
axial--symmetric top with moments of inertia $I_{\perp}$ and
$I_{\parallel}$. Within this regularization the evolution operator
representing Wilson loops has been replaced by a path integral
over dynamical variables of this axial--symmetric top which correspond to gauge degrees of freedom of the non--Abelian gauge field.  The regularized expression of the evolution
operator has been obtained in the limit $I_{\perp}, I_{\parallel} \to
0$. The moments of inertia have been used as parameters
like $\hbar \to 0$. Unfortunately, as we show in Sect.\,6
the limit $I_{\perp}, I_{\parallel} \to 0$ has been evaluated
incorrectly.

\section{The $SU(N)$ extension}
\setcounter{equation}{0}

The extension of the path integral representation given in
Eq.(\ref{label2.24}) to $SU(N)$ is rather straightforward and reduces
to the evaluation of the character of the matrix $U^{\Omega}_r(C_{x_k
x_{k-1}})$ in the irreducible representation $r$ of $SU(N)$. The
character can be given by [12]
\begin{eqnarray}\label{label3.1}
\chi[U^{\Omega}_r(C_{x_k x_{k-1}})] &=& {\rm tr}(e^{\textstyle 
i\sum^{N-1}_{{\ell} = 1} H_{\ell}\Phi_{\ell}[C_{x_k x_{k-1}};
A^{\Omega}]}) = \nonumber\\
&=& \sum_{\vec{m}_r}\gamma_{\,\vec{m}_r}\,e^{\textstyle 
i\,\vec{m}_r\cdot {\vec{\Phi}}[C_{x_k x_{k-1}}; A^{\Omega}]},
\end{eqnarray}
where $H_{\ell}\,({\ell} = 1,\ldots,N-1)$ are diagonal $d_r\times d_r$
traceless matrices realizing the representation of the Cartan
subalgebra, i.e.  $[H_{\ell},H_{\ell'}]=0$, of the generators of the
$SU(N)$ [12].  The sum runs over all the weights $\vec{m}_r
=(m_{r\,1},\ldots,m_{r\,N-1})$ of the irreducible
representation $r$ and $\gamma_{\,\vec{m}_r}$ is the
multiplicity of the weight $\vec{m}_r$ and
$\sum_{\vec{m}_r}\gamma_{\,\vec{m}_r} = d_r$.  The
components of the vector ${\vec{\Phi}}[C_{x_k x_{k-1}}; A^{\Omega}]$
are defined by
\begin{eqnarray}\label{label3.2}
\Phi_{\ell}[C_{x_k x_{k-1}}; A^{\Omega}] = g\int_{C_{x_k
x_{k-1}}}\!\!\! {\varphi}_{\ell}\,[\omega(x)],
\end{eqnarray}
where we have introduced the notation $\omega(x) = t^a\omega^a(x) =
dz\cdot A^{\Omega}(x)$. The functions ${\varphi}_{\ell}\,[\omega(x)]$
are proportional to the roots of the equation ${\rm det}\Big[\omega(x)
- \lambda\Big]=0$.

The path integral representation of Wilson loops
defined for the irreducible representation $r$ of $SU(N)$ reads
\begin{eqnarray}\label{label3.3}
\hspace{-0.3in}W_r(C) =\frac{1}{d^2_r} 
\int \prod_{x\in C} D\Omega_r(x) \sum_{\left\{ \vec{m}_r(x) \right\} } \,d_r\,\gamma_{\,\vec{m}_r(x)}\, e^{\textstyle
i\,g\,\oint_{C}\vec{m}_r(x)\cdot \vec{\varphi}\,[\omega(x)]}.
\end{eqnarray}
Let us consider in more details the path integral representation of
Wilson loops defined for the fundamental representation
$\underline{3}$ of $SU(3)$. The character
$\chi_{\underline{3}}[U^{\Omega}_{\underline{3}}(C)]$ is defined as
\begin{eqnarray}\label{label3.4}
\hspace{-0.3in}&&\chi_{\underline{3}}[U^{\Omega}_{\underline{3}}(C)]
= {\rm tr}\Big(e^{\textstyle  iH_1\Phi_1[C; A^{\Omega}] + iH_2\Phi_2[C;
A^{\Omega}]}\,\Big) = e^{\textstyle - i\Phi_2[C; A^{\Omega}]/3} \nonumber\\
\hspace{-0.3in}&&+ e^{\textstyle i\Phi_1[C;
A^{\Omega}]/2\sqrt{3}}\,e^{\textstyle i\Phi_2[C;
A^{\Omega}]/6} + e^{\textstyle -
i\Phi_1[C; A^{\Omega}]/2\sqrt{3}}\,e^{\textstyle  i\Phi_2[C;
A^{\Omega}]/6},
\end{eqnarray}
where $H_1 = t^3/\sqrt{3}$ and $H_2 = t^8/\sqrt{3}$ [12]. For the
representation $\underline{3}$ of $SU(3)$ the equation
${\rm det}[\omega - \lambda]=0$ takes the form
\begin{eqnarray}\label{label3.5}
\lambda^3 - \lambda\,\frac{1}{2}\,{\rm tr}\,\omega^2(x) - {\rm
det}\,\omega(x) = 0.
\end{eqnarray}
The roots of Eq.(\ref{label3.5}) read
\begin{eqnarray}\label{label3.6}
\lambda^{(1)}&=& - \frac{1}{\sqrt{6}}\,\sqrt{{\rm
tr}\,\omega^2(x)}\,\cos\Bigg(\frac{1}{3}\,{\rm
arccos}\sqrt{2\,{\rm det}\Bigg[\displaystyle 1 +
12\,\frac{t^a{\rm tr}(t^a \omega^2(x))}{{\rm
tr}\,\omega^2(x)}\Bigg]}\,\Bigg) \nonumber\\ &&-
\frac{1}{\sqrt{2}}\,\sqrt{{\rm
tr}\,\omega^2(x)}\,\sin\Bigg(\frac{1}{3}\,{\rm
arccos}\sqrt{2\,{\rm det}\Bigg[\displaystyle 1 +
12\,\frac{t^a{\rm tr}(t^a \omega^2(x))}{{\rm
tr}\,\omega^2(x)}\Bigg]}\,\Bigg) ,\nonumber\\ \lambda^{(2)}&=& -
\frac{1}{\sqrt{6}}\,\sqrt{{\rm
tr}\,\omega^2(x)}\,\cos\Bigg(\frac{1}{3}\,{\rm
arccos}\sqrt{2\,{\rm det}\Bigg[\displaystyle 1 +
12\,\frac{t^a{\rm tr}(t^a \omega^2(x))}{{\rm
tr}\,\omega^2(x)}\Bigg]}\,\Bigg)  \nonumber\\ &&+
\frac{1}{\sqrt{2}}\,\sqrt{{\rm
tr}\,\omega^2(x)}\,\sin\Bigg(\frac{1}{3}\,{\rm
arccos}\sqrt{2\,{\rm det}\Bigg[\displaystyle 1 +
12\,\frac{t^a{\rm tr}(t^a \omega^2(x))}{{\rm
tr}\,\omega^2(x)}\Bigg]}\,\Bigg) ,\nonumber\\
\lambda^{(3)}&=&~\,\sqrt{\frac{2}{3}}\,\sqrt{{\rm
tr}\,\omega^2(x)}\,\cos\Bigg(\frac{1}{3}\,{\rm
arccos}\sqrt{2\,{\rm det}\Bigg[\displaystyle 1 +
12\,\frac{t^a{\rm tr}(t^a \omega^2(x))}{{\rm
tr}\,\omega^2(x)}\Bigg]}\,\Bigg) .\nonumber\\ &&
\end{eqnarray}
In terms of the roots $\lambda^{(i)}\,(i =1,2,3)$ the phases
$\Phi_{1,2}[C; A^{\Omega}]$ are defined as
\begin{eqnarray}\label{label3.7}
\hspace{-0.3in}&&\Phi_1[C;
A^{\Omega}]= -g\sqrt{6}\oint_C\sqrt{{\rm
tr}\,\omega^2(x)}\sin\Bigg(\frac{1}{3}\,{\rm
arccos}\sqrt{2\,{\rm det}\Bigg[\displaystyle 1 +
12\frac{t^a{\rm tr}(t^a \omega^2(x))}{{\rm
tr}\,\omega^2(x)}\Bigg]}\,\Bigg),\nonumber\\
\hspace{-0.3in}&&\Phi_2[C;
A^{\Omega}]= - g\sqrt{6}\oint_C\sqrt{{\rm
tr}\,\omega^2(x)}\cos\Bigg(\frac{1}{3}{\rm
arccos}\sqrt{2\,{\rm det}\Bigg[\displaystyle 1 +
12\frac{t^a{\rm tr}(t^a \omega^2(x))}{{\rm
tr}\,\omega^2(x)}\Bigg]}\,\Bigg),
\end{eqnarray}
where ${\rm tr}\,\omega^2(x) =
\frac{1}{2}\,g_{\mu\nu}[A^{\Omega}](x)\,dx_{\mu}dx_{\nu}$. Thus, in
the fundamental representation $\underline{3}$ the path integral representation for Wilson loops reads
\begin{eqnarray}\label{label3.8}
\hspace{-0.3in}W_{\underline{3}}(C)&=&\frac{1}{9}\int \prod_{x\in
C} \left[D \Omega_{\underline{3}}(x)\times\,3 \right] \, \Big(
e^{\textstyle i\Phi_1[C; A^{\Omega}]/2\sqrt{3}}e^{\textstyle
i\Phi_2[C; A^{\Omega}]/6} \nonumber\\
\hspace{-0.3in}&& + e^{\textstyle -
i\Phi_1[C; A^{\Omega}]/2\sqrt{3}}\,e^{\textstyle  i\Phi_2[C;
A^{\Omega}]/6} + e^{\textstyle - i\Phi_2[C; A^{\Omega}]/3}\Big),
\end{eqnarray}
where the phases $\Phi_{1,2}[C; A^{\Omega}]$ are given by
Eq.(\ref{label3.7}).

\section{Wilson loop for pure gauge field}
\setcounter{equation}{0}

As has been pointed out in Ref.[3] the path integral over gauge
degrees of freedom representing Wilson loops {\it is not of the
Feynman type, therefore, it depends explicitly on how one
``understands'' it, i.e. how it is discretized and regularized}. We
would like to emphasize that the {\it regularization procedure}
applied in Ref.[3] has led to an expression for Wilson loops which
supports the hypothesis of Maximal Abelian Projection [14]. According
to this hypothesis only Abelian degrees of freedom of non--Abelian
gauge fields are responsible for confinement. This is to full extent a
dynamical hypothesis. It is quite obvious that such a dynamical
hypothesis cannot be derived only by means of a regularization
procedure.

In order to show that the problem touched in this paper is not of
marginal interest and to check if path integral expressions that look differently superficially could actually compute the same number we evaluate below
explicitly the path integrals representing Wilson loop for a pure
$SU(2)$ gauge field. As has been stated in Ref.[3] for Wilson loops
$C$ a gauge field {\it along a given curve can be always written as a
``pure gauge''} and the derivation of the path integral representation
for Wilson loops can be provided for the gauge field taken {\it
without loss of generality in the ``pure gauge'' form}. We would like
to show that for the pure $SU(2)$ gauge field the path integral
representation for Wilson loops suggested in Ref.[3] fails for a
correct description of Wilson loops. Since a pure gauge field is
equivalent to a zero gauge field Wilson loops should be unity.

Of course, any correct path integral representation for Wilson
loops should lead to the same result. The evaluation of Wilson loops
within the path integral representation Eq.(\ref{label2.8}) is rather
trivial and transparent. Indeed, we have not corrupted the starting
expression for Wilson loops (\ref{label2.1}) by any artificial
regularization. Thereby, the general formula (\ref{label2.8})
evaluated through the discretization given by Eqs.(\ref{label2.7}) and
(\ref{label2.6}) is completely identical to the original expression
(\ref{label2.1}). The former gives a unit value for Wilson loops
defined for an arbitrary contour $C$ and an irreducible representation
$J$ of $SU(2)$: $W_J(C)=1$. 

Let us focus now on the path integral representation suggested in
Ref.[3]
\begin{eqnarray}\label{label4.1}
W_J(C)=\int \prod_{x\in C} D \Omega(x)\,e^{\textstyle 2 i J g 
\oint_{C}dx_{\mu}\,{\rm tr}[t^3 A^{\Omega}_{\mu}(x)]},
\end{eqnarray}
where all matrices are taken in the irreducible representation $J$.
Following the discretization suggested in Ref.[3] we arrive at the
expression
\begin{eqnarray}\label{label4.2}
W_J(C)= \lim_{n\to \infty}\prod^{n}_{k=1}\int D
\Omega(x_k)\,e^{\textstyle 2 i J g\int_{C_{x_{k+1} x_k}}dx_{\mu}\,{\rm
tr}[t^3 A^{\Omega}_{\mu}(x)]}.
\end{eqnarray}
Setting $A_{\mu}(x)=\partial_{\mu}U(x) U^{\dagger}(x)/ig$ we get
\begin{eqnarray}\label{label4.3}
A^{\Omega}_{\mu}(x)=\frac{1}{ig}\partial_{\mu}(\Omega(x)U(x))(\Omega(x)
U(x))^{\dagger}.
\end{eqnarray}
By a gauge transformation $\Omega(x)U(x) \to
\Omega(x)$ we reduce Eq.(\ref{label4.1}) to the form 
\begin{eqnarray}\label{label4.4}
W_J(C)=\int \prod_{x\in C} D \Omega(x)\,e^{\textstyle 
2 J\oint_{C}dx_{\mu}\,{\rm tr}[t^3\partial_{\mu}\Omega(x)
\Omega^{\dagger}(x)]}.
\end{eqnarray}
For simplicity we consider Wilson loops in the fundamental
representation of $SU(2)$,
$W_{1/2}(C)$. The result can be generalized to any irreducible
representation $J$.

For the evaluation of the path integral Eq.(\ref{label4.4}) it is
convenient to use a standard $s$--parameterization of Wilson loops
$C$ [2]: $x_{\mu} \to x_{\mu}(s)$, with $s \in
[0,1]$ and $x_{\mu}(0) = x_{\mu}(1) = x_{\mu}$.

The Wilson loop (\ref{label4.4}) reads in the $s$--parameterization
\begin{eqnarray}\label{label4.5}
W_{1/2}(C)=\int \prod_{0 \le s \le 1} D \Omega(s)\,\exp  \int\limits^1_0
ds\,{\rm tr}\Bigg[t^3\frac{d\Omega(s)}{ds}\Omega^{\dagger}(s)\Bigg].
\end{eqnarray}
The discretized form of the path integral (\ref{label4.5}) is given by 
\begin{eqnarray}\label{label4.6}
&&W_J(C) = \lim_{n \to\infty}\int \prod^n_{k = 1} D \Omega_k\,\exp 
\Delta s_{k+1,k}\,{\rm
tr}\Bigg[t^3\frac{\Omega_{k+1}-\Omega_{k}}{\Delta
s_{k+1,k}}\Omega^{\dagger}_k\Bigg]=\nonumber\\ 
&& = \lim_{n
\to\infty}\int \prod^n_{k = 1} D \Omega_k\,e^{\textstyle {\rm
tr}[t^3\Omega_{k+1}\Omega^{\dagger}_k]} = \lim_{n\to \infty}\int
\ldots \int D\Omega_n D\Omega_{n-1} D\Omega_{n-2}\ldots
D\Omega_1\,\nonumber\\ &&\times\, e^{\textstyle {\rm tr}[t^3\Omega_n
\Omega^{\dagger}_{n-1}]}\, e^{\textstyle 2J{\rm
tr}[t^3\Omega_{n-1}\Omega^{\dagger}_{n-2}]} \ldots e^{\textstyle  {\rm
tr}[t^3\Omega_2 \Omega^{\dagger}_1]} \, e^{\textstyle {\rm
tr}[t^3\Omega_1 \Omega^{\dagger}_n]},
\end{eqnarray}
where $\Omega_{n+1} = \Omega_{1}$.

For the subsequent integration over $\Omega_k$ we suggest to use a
formula of Ref.[15] modified for our case
\begin{eqnarray}\label{label4.7}
\int D \Omega\,e^{\textstyle z{\rm tr}[t^3 A\Omega^{\dagger} + Bt^3\Omega]} =
\sum_{j} \frac{a^2_j(z)}{2j+1}\,\chi_j[(t^3)^2AB],
\end{eqnarray}
where the coefficients $a_j(z)$ are defined by the expansion [15]
\begin{eqnarray}\label{label4.8}
e^{\textstyle z{\rm tr}[t^3U]} = \sum_{j} a_j(z)\,\chi_j[t^3U]. 
\end{eqnarray}
In the particular case $z = 2J$ and  for the fundamental representation $J
= 1/2$ we have $z = 1$.  The trace ${\rm tr}[t^3U]$ in the exponent of
the l.h.s. of Eq.(\ref{label4.8}) should be evaluated for the
fundamental representation of $SU(2)$.  By virtue of the
orthogonality relation for characters [9,10,15]
\begin{eqnarray}\label{label4.9}
\int D U \,\chi_j[A U^{\dagger}]\,\chi_{j^{\prime}}[U B] =
\frac{\delta_{j j^{\prime}}}{2j+1}\,\chi_j[AB],
\end{eqnarray}
where $D U$ is the Haar measure for the $SU(2)$ group, the
coefficients $a_j(z)$ for $j \not= 0$ can be determined by [15]
\begin{eqnarray}\label{label4.10}
a_j(z) = \frac{3}{j(j+1)}\int D U\,\chi_j[t^3U^{\dagger}]\,e^{\textstyle
z{\rm tr}[t^3U]}.
\end{eqnarray}
We have used here that $\chi_j[(t^3)^2] = j(j+1)(2j+1)/3$. The
coefficient $a_0(z)$ is defined by
\begin{eqnarray}\label{label4.11}
a_0(z) = \int D U\,e^{\textstyle z{\rm tr}[t^3U]}.
\end{eqnarray}
The coefficients $a_j(z)$ obey a completeness condition. For its
derivation we notice that ${\rm tr}[t^3U^{\dagger}]= - {\rm
tr}[t^3U]$ which can be easily seen from the standard parameterization
of the matrix $U$ in terms of an angle $\varphi$ and a unit vector
$\vec{n}$ (see Eq.(3.96) of Ref.[10])
\begin{eqnarray}\label{label4.12}
U &=& e^{\displaystyle
+ i\varphi\,\vec{n}\cdot\vec{\tau}/2} = \cos\frac{\varphi}{2}+
i(\vec{n}\cdot\vec{\tau}\,)\sin\frac{\varphi}{2},\nonumber\\
U^{\dagger} &=& e^{\displaystyle -
i\varphi\,\vec{n}\cdot\vec{\tau}/2} = \cos\frac{\varphi}{2}-
i(\vec{n}\cdot\vec{\tau}\,)\sin\frac{\varphi}{2},
\end{eqnarray}
The expansion of $e^{\textstyle z{\rm
tr}[t^3U^{\dagger}]}$ we represent as follows
\begin{eqnarray}\label{label4.13}
e^{\textstyle z{\rm tr}[t^3U^{\dagger}]} = \sum_{j}
b_j(z)\,\chi_j[t^3U^{\dagger}].
\end{eqnarray}
Let us show that $b_j(z) = a_j(z)$. Using the orthogonality relation
(\ref{label4.9}) we obtain
\begin{eqnarray}\label{label4.14}
b_j(z)&=& \frac{3}{j(j+1)}\int D U\,\chi_j[t^3U]\,e^{\textstyle
z{\rm tr}[t^3U^{\dagger}]},\, j\not= 0,\nonumber\\
b_0(z)&=& \int D U\,e^{\textstyle
z{\rm tr}[t^3U^{\dagger}]}.
\end{eqnarray}
Then, making the change $U^{\dagger} \to U$ we get $b_j(z) = a_j(z)$ by virtue
of the invariance of the Haar measure $D U^{\dagger} = D U$.

Thus, taking the product of the expansions (\ref{label4.8}) and
(\ref{label4.13}) with $b_j(z) = a_j(z)$ and integrating over $U$ we get
\begin{eqnarray}\label{label4.15}
&&\int D U\,e^{\textstyle z{\rm tr}[t^3U]}\,e^{\textstyle z{\rm
tr}[t^3U^{\dagger}]} = \sum_{j}\sum_{j'} a_j(z)\,a_{j'}(z)\int D
U\,\chi_j[t^3U^{\dagger}]\,\chi_{j'}[t^3U] =\nonumber\\
&&=\sum_{j}\frac{a^2_j(z)}{2j+1}\,\chi_j[(t^3)^2] = a^2_0(z) + 
\sum_{j > 0}\frac{1}{3}\,j(j+1)\,a^2_j(z).
\end{eqnarray}
The l.h.s. of Eq.(\ref{label4.15}) is equal to unity due to the
relation ${\rm tr}[t^3U^{\dagger}]= - {\rm tr}[t^3U]$ and the
normalization of the Haar measure $\int D U = 1$. Therefore, the completeness condition for the coefficients $a_j(z)$ reads
\begin{eqnarray}\label{label4.16}
a^2_0(z) + \sum_{j > 0}\frac{1}{3}\,j(j+1)\,a^2_j(z) = 1.
\end{eqnarray}
The coefficient $a_0(z)$ we evaluate below. The evaluation of
coefficients $a_j(z)$ for an arbitrary $j$ is given in the 
Appendix. For the evaluation of $a_0(z)$ one can use, for example,
the standard parameterization (\ref{label4.12}) and the definition
of the Haar measure $DU$ (see Eq.(3.97) of Ref.[10])
\begin{eqnarray}\label{label4.17}
DU = \frac{1}{4\pi^2}\,d\Omega_{\vec{n}}\,d\varphi\,\sin^2\frac{\varphi}{2},
\end{eqnarray}
where $d\Omega_{\vec{n}}$ is the uniform measure on the unit sphere
$S^2$ [9]. As a result for  $a_0(z)$ we obtain
\begin{eqnarray}\label{label4.18}
a_0(z) =\int D U\,e^{\textstyle z{\rm tr}[t^3U]} = 2 J_1(z)/z,
\end{eqnarray}
where $J_1(z)$ is a Bessel function [16]. In the particular case, $z
=1$, we get $a_0(1) = a_0 = 2 J_1(1) = 0.88$ [16].

For the integration over $\Omega_k$ we suppose, first, that $n$ is an
even number. Then, integrating over $\Omega_{n-1}$,
$\Omega_{n-3}$,$\ldots$, $\Omega_{1}$ we obtain
\begin{eqnarray}\label{label4.19}
\hspace{-0.5in}&&W_{1/2}(C)= \lim_{n\to \infty}\int \ldots \int D\Omega_n
D\Omega_{n-2} D\Omega_{n-4}\ldots D\Omega_2\,\nonumber\\
\hspace{-0.5in}&&\times\,\sum_{j_{n-1}} (2j_{n-1} +
1)\,\Bigg[\frac{a_{j_{n-1}}}{2j_{n-1}+1}\Bigg]^2
\,\chi_{j_{n-1}}[(t^3)^2\Omega_n \Omega^{\dagger}_{n-2}]\nonumber\\
\hspace{-0.5in}&&\times\,\sum_{j_{n-3}} (2j_{n-3} +
1)\,\Bigg[\frac{a_{j_{n-3}}}{2j_{n-3}+1}\Bigg]^2 \,\chi_{j_{n-3}}[(t^3)^2
\Omega_{n-2}\Omega^{\dagger}_{n-4}]\,\ldots \nonumber\\ 
\hspace{-0.5in}&&\times\,\sum_{j_4} (2j_4 +
1)\,\Bigg[\frac{a_{j_4}}{2j_4+1}\Bigg]^2 \,\chi_{j_4}[(t^3)^2\Omega_4
\Omega^{\dagger}_2]\,\nonumber\\
\hspace{-0.5in}&&\times\,\sum_{j_2} (2j_2 +
1)\,\Bigg[\frac{a_{j_2}}{2j_2+1}\Bigg]^2
\,\chi_{j_2}[(t^3)^2\Omega_2\Omega^{\dagger}_n],
\end{eqnarray}
where we have denoted $a_j(1) = a_j$.

After the integration over $\Omega_{n}$, $\Omega_{n-2}$,$\ldots$,
$\Omega_{2}$ we arrive at the expression
\begin{eqnarray}\label{label4.20}
W_{1/2}(C) &=& \lim_{n\to \infty}\sum_{j} (2j +
1)\,\Bigg[\frac{a_j}{2j+1}\Bigg]^n\,\chi_j[(t^3)^n] =\nonumber\\
&=&\lim_{n\to \infty}\Big(a^n_0 +\sum_{j > 0} (2j +
1)\,\Bigg[\frac{a_j}{2j+1}\Bigg]^n\,\chi_j[(t^3)^n]\Big) =\nonumber\\
&=& \lim_{n\to \infty}\sum_{j > 0} (2j +
1)\,\Bigg[\frac{a_j}{2j+1}\Bigg]^n\,\chi_j[(t^3)^n],
\end{eqnarray}
where we have used that $\lim_{n\to \infty}a^n_0 = 0$ by virtue of the
relation $a_0 = 2 J_1(1) = 0.88 < 1$ given by Eq.(\ref{label4.18}). 

The expression (\ref{label4.20}) is valid too if $n$ is an odd
number. However, in this case $\chi_j[(t^3)^n] = 0$ and we get
immediately $W_{1/2}(C) = 0$.

In order to estimate $\chi_j[(t^3)^n]$ for $n$ an even number and at $n
\to \infty$ we suggest to apply  the following procedure
\begin{eqnarray}\label{label4.21}
&&\chi_j[(t^3)^n] = \sum^{j}_{m = -j} m^n =\nonumber\\ &&=
\frac{\Gamma(n+1)}{2\pi i}\int\limits^{i\infty}_{-i\infty}\sum^{j}_{m
= -j}e^{\textstyle m s}\frac{d s}{s^{n+1}} =
\frac{\Gamma(n+1)}{2\pi
i}\int\limits^{i\infty}_{-i\infty}\frac{\displaystyle{\rm
sh}(2j+1)\frac{s}{2}}{\displaystyle{\rm sh}\,\frac{s}{2}}\frac{d
s}{s^{n+1}}.
\end{eqnarray}
As $n \to \infty$, we can evaluate the integral over $s$ by using the
saddle--point approach and get $\chi_j[(t^3)^n] \simeq j^n$. 

The Wilson loop is then defined by
\begin{eqnarray}\label{label4.22}
W_{1/2}(C)= \lim_{n\to \infty}\sum_{j > 0} (2j +
1)\,\Bigg[\frac{j\,a_j}{2j+1}\Bigg]^n = \lim_{n\to \infty}\sum_{j > 0}
(2j + 1)\,\exp \Bigg[n\,{\ell n}\Bigg|\frac{j\,a_j}{2j+1}\Bigg|\Bigg].
\end{eqnarray}
By using the completeness condition for  the coefficients $a_j$
given by Eq.(\ref{label4.16}) we obtain the constraint 
\begin{eqnarray}\label{label4.23}
\Bigg|\frac{j\,a_j}{2j+1}\Bigg| < \sqrt{3(1-a^2_0)}\sqrt{\frac{
j}{j+1}}\,\frac{1}{2j+1} < \sqrt{\frac{
j}{j+1}}\,\frac{1}{2j+1} < 1.
\end{eqnarray}
This proves that the Wilson loop $W_{1/2}(C)$ vanishes in the limit $n
\to \infty$, $W_{1/2}(C) = 0$.

Thus, the Wilson loop $W_{1/2}(C)$ for an arbitrary contour $C$ and a
pure gauge field represented by the path integral derived in Ref.[3]
vanishes, instead of being equal to unity,
$W_{1/2}(C) = 1$. This shows that the path integral representation
suggested in Ref.[3] fails for the correct
description of Wilson loops.

\section{Wilson loop for $Z(2)$ center vortices}
\setcounter{equation}{0}

In this Section we evaluate explicitly the path
 integral (\ref{label4.1}) for Wilson loops pierced by a $Z(2)$
 center vortex with spatial  azimuthal symmetry.  Some problems of
 $Z(2)$ center vortices with spatial  azimuthal symmetry have been
 analysed by Diakonov in his recent publication [17] for the gauge group $SU(2)$.  In this system the main dynamical variable is the
 azimuthal component of the non--Abelian gauge field
 $A^a_{\phi}(\rho)\,$ ($a = 1,2,3$) depending only on $\rho$, the
 radius in the transversal plane.  For a circular Wilson loop in the
 irreducible representation $J$ one gets 
\begin{eqnarray}\label{label5.1}
W_J(\rho) = \frac{1}{2J+1}\sum^{J}_{m = - J}e^{\textstyle i 2\pi m 
\mu(\rho)}= \frac{1}{2J+1}\frac{\sin[(2J+1)\pi
\mu(\rho)]}{\sin[\pi \mu(\rho)]},
\end{eqnarray}
where $\mu(\rho) = \rho \sqrt{A^a_{\phi}(\rho) A^a_{\phi}(\rho)}$. The
gauge coupling constant $g$ is included in the definition of the gauge
field. For Wilson loops in the fundamental representation $J=1/2$
we have
\begin{eqnarray}\label{label5.2}
W_{1/2}(\rho) = \cos[\pi \mu(\rho)].
\end{eqnarray}
In the case of $Z(2)$ center vortices with spatial azimuthal symmetry
and for the fundamental representation of $SU(2)$ Eq.(\ref{label4.2}) takes the form
\begin{eqnarray}\label{label5.3}
W_{1/2}(\rho)&=& \lim_{n\to \infty}\prod^{n}_{k=1}\int D
\Omega_k\,e^{\textstyle {\rm tr}[t^3(i2\pi\rho/n)\Omega_{k+1}
A_{\phi}(\rho)\Omega^{\dagger}_k +
t^3 \Omega_{k+1}\Omega^{\dagger}_k]}= \nonumber\\ 
&&=\lim_{n\to \infty}\int \ldots 
\int D\Omega_n D\Omega_{n-1} D\Omega_{n-2}\ldots
D\Omega_1 \nonumber\\
&&\times\,e^{\textstyle 
\,{\rm tr}[t^3(i2\pi\rho/n)\Omega_n  A_{\phi}(\rho)\Omega^{\dagger}_{n-1}
 + t^3\Omega_n \Omega^{\dagger}_{n-1}]}\nonumber\\
&&\times\,e^{\textstyle {\rm tr}[t^3(i2\pi\rho/n)\Omega_{n-1}
 A_{\phi}(\rho)\Omega^{\dagger}_{n-2} + t^3\Omega_{n-1}
\Omega^{\dagger}_{n-2}]}\,\ldots \nonumber\\
&&\times\,e^{\textstyle 
\,{\rm tr}[t^3(i2\pi\rho/n)\Omega_2 A_{\phi}(\rho)\Omega^{\dagger}_1 + 
t^3\Omega_2 \Omega^{\dagger}_1]}\nonumber\\
&&\times\,e^{\textstyle {\rm tr}[t^3(i2\pi\rho/n)\Omega_1  
A_{\phi}(\rho)\Omega^{\dagger}_n + t^3\Omega_1\Omega^{\dagger}_n]},
\end{eqnarray}
where we have used $C_{x_{k+1} x_k} = 2\pi\rho/n$, $\Omega(x_k) = \Omega_k$ and $\Omega_{n+1} = \Omega_1$.

For the subsequent evaluation it is convenient to introduce the matrix
\begin{eqnarray}\label{label5.4}
Q(A_{\phi}) = \Big(1 + i\,\frac{2\pi}{n}\,\rho\, A_{\phi}(\rho)\Big).
\end{eqnarray}
In terms of $Q(A_{\phi})$ the path integral (\ref{label5.3}) reads
\begin{eqnarray}\label{label5.5}
&&W_{1/2}(\rho) = \lim_{n\to \infty}\int \ldots \int D\Omega_n
D\Omega_{n-1} D\Omega_{n-2}\ldots D\Omega_1\,e^{\textstyle {\rm
tr}[t^3\Omega_n Q(A_{\phi})\Omega^{\dagger}_{n-1}]}\, \nonumber\\
&&\times\,e^{\textstyle {\rm tr}[t^3\Omega_{n-1}Q(A_{\phi})
\Omega^{\dagger}_{n-2}]}\,\ldots \,e^{\textstyle {\rm tr}[ t^3\Omega_2
Q(A_{\phi})\Omega^{\dagger}_1]}\,e^{\textstyle \,{\rm tr}[t^3\Omega_1
Q(A_{\phi})\Omega^{\dagger}_n]},
\end{eqnarray}
The integration over $\Omega_k$ we carry out with the help of
Eq.(\ref{label4.7}) taken in the from
\begin{eqnarray}\label{label5.6}
&&\int D \Omega_k\,e^{\textstyle {\rm tr}[t^3\Omega_{k+1}
Q(A_{\phi})\Omega^{\dagger}_k + Q(A_{\phi})
\Omega^{\dagger}_{k-1}t^3\Omega_k]} = \nonumber\\
&&=\sum_{j}
\frac{a^2_j}{2j+1}\,\chi_j[(t^3)^2\Omega_{k+1}
Q^2(A_{\phi})\Omega^{\dagger}_{k-1}]
\end{eqnarray}
and the orthogonality relation (\ref{label4.9}). The coefficients
$a_j$ obey the completeness condition (\ref{label4.16}) with the
constraint (\ref{label4.23}). 

The number $n$ may be both even or odd. Let $n$ be an even number,
then integrating over $\Omega_{n-1}$, $\Omega_{n-3}$,$\ldots$,
$\Omega_1$ by using Eq.(\ref{label5.6}) we obtain
\begin{eqnarray}\label{label5.7}
\hspace{-0.3in}&&W_{1/2}(\rho) = \lim_{n\to \infty} \int \ldots \int
D\Omega_n D\Omega_{n-2} D\Omega_{n-4}\ldots D\Omega_2\,\nonumber\\
\hspace{-0.5in}&&\times\,\sum_{j_{n-1}} (2j_{n-1} +
1)\,\Bigg[\frac{a_{j_{n-1}}}{2j_{n-1}+1}\Bigg]^2
\,\chi_{j_{n-1}}[(t^3)^2\Omega_n Q^2(A_{\phi})
\Omega^{\dagger}_{n-2}]\nonumber\\
\hspace{-0.5in}&&\times\,\sum_{j_{n-3}} (2j_{n-3} +
1)\,\Bigg[\frac{a_{j_{n-3}}}{2j_{n-3}+1}\Bigg]^2
\,\chi_{j_{n-3}}[(t^3)^2
\Omega_{n-2}Q^2(A_{\phi})\Omega^{\dagger}_{n-4}]\,\ldots \nonumber\\
\hspace{-0.5in}&&\times\,\sum_{j_4} (2j_4 +
1)\,\Bigg[\frac{a_{j_4}}{2j_4+1}\Bigg]^2 \,\chi_{j_4}[(t^3)^2\Omega_4
Q^2(A_{\phi}) \Omega^{\dagger}_2]\,\nonumber\\
\hspace{-0.5in}&&\times\,\sum_{j_2} (2j_2 +
1)\,\Bigg[\frac{a_{j_2}}{2j_2+1}\Bigg]^2
\,\chi_{j_2}[(t^3)^2\Omega_2 Q^2(A_{\phi}) \Omega^{\dagger}_n].
\end{eqnarray}
The integration over $\Omega_n$,  $\Omega_{n-2}$, $\ldots$, $\Omega_2$ gives
\begin{eqnarray}\label{label5.8}
W_J(\rho) &=& \lim_{n\to \infty}\sum_{j} (2j +
1)\,\Bigg[\frac{a_j}{2j+1}\Bigg]^n\,\int
D\Omega_n\,\chi_j[(t^3)^n\Omega_n Q^n(A_{\phi}) \Omega^{\dagger}_n] =
\nonumber\\ &=& \lim_{n\to \infty}\Big(a^n_0 + \sum_{j > 0}
\Bigg[\frac{a_j}{2j+1}\Bigg]^n\,
\chi_j[(t^3)^n]\,\chi_j[Q^n(A_{\phi})]\Big)= \nonumber\\ &=&
\lim_{n\to \infty} \sum_{j > 0} \Bigg[\frac{a_j}{2j+1}\Bigg]^n\,
\chi_j[(t^3)^n]\,\chi_j[Q^n(A_{\phi})],
\end{eqnarray}
where we have set $\lim_{n\to \infty}a^n_0= 0$.

The integration over $\Omega_n$ we have carried out by means of
Eq.(\ref{label2.3}).  One can easily show that Eq.(\ref{label5.8}) is
also valid for odd $n$, as well as Eq.(\ref{label4.12}). In this
case due to the relation $\chi_j[(t^3)^n] =0$ we obtain again
$W_{1/2}(\rho) = 0$.

For even $n$ we should use the relation $\chi_j[(t^3)^n]
\simeq j^n$ at $n \to \infty$ which follows from Eq.(\ref{label4.13}). This reduces the r.h.s. of Eq.(\ref{label5.7}) to the form
\begin{eqnarray}\label{label5.9}
W_{1/2}(\rho)  = \lim_{n\to \infty}\sum_{j > 0}
\Bigg[\frac{j a_j}{2j+1}\Bigg]^n\,\chi_j[Q^n(A_{\phi})],
\end{eqnarray}
The evaluation of the character $\chi_j[Q^n(A_{\phi})]$ for $n \to
\infty$ runs as follows
\begin{eqnarray}\label{label5.10}
\chi_j[Q^n(A_{\phi})] &=& \chi_j\Big[\Big(1 +
i\,\frac{2\pi}{n}\,\rho\, A_{\phi}(\rho)\Big)^n\Big] \simeq
\chi_j\Big[e^{\textstyle i 2\pi \rho A_{\phi}(\rho)}\Big] =
\nonumber\\ &=&\frac{\sin[(2j+1)\pi \mu(\rho)]}{\sin[\pi
\mu(\rho)]}.
\end{eqnarray}
Substituting Eq.(\ref{label5.10}) in Eq.(\ref{label5.9}) we obtain
\begin{eqnarray}\label{label5.11}
W_{1/2}(\rho)= \lim_{n\to \infty}\sum_{j > 0}
\Bigg[\frac{j a_j}{2j+1}\Bigg]^n\,\frac{\sin[(2j+1)\pi
\mu(\rho)]}{\sin[\pi \mu(\rho)]}.
\end{eqnarray}
Due to the constraint Eq.(\ref{label4.23}) the Wilson loop vanishes in
the limit $n \to \infty$, $W_{1/2}(\rho) = 0$. Thus, we have shown
that the the path integral for Wilson loops suggested in Ref.[3] gives zero for a field configuration with a  $Z(2)$ center vortex,
$W_{1/2}(\rho) = 0$, instead of the correct result $W_{1/2}(\rho) =
\cos\pi\mu(\rho)$, Eq.(\ref{label5.2}).

We hope that the examples considered in Sect.4 and 5 demonstrate
that the path integral
representation for Wilson loops derived in Ref.[3] is
erroneous. Nevertheless, in Sect.\,6 we evaluate explicitly
 the regularized evolution operator $Z_{\rm Reg}(R_2,R_1)$ suggested by Diakonov and Petrov for the representation of the
Wilson loop in Ref.[3]. We show that this regularized evolution operator
$Z_{\rm Reg}(R_2,R_1)$ has been evaluated incorrectly in Ref.[3]. The
correct evaluation gives $Z_{\rm Reg}(R_2,R_1) = 0$ which agrees with
our results obtained above.

\section{Path integral for the evolution operator $Z(R_2,R_1)$ } 
\setcounter{equation}{0}

As has been suggested in Ref.[3] the functional  $Z(R_2,R_1)$
defined by (see Eq.(8) of Ref.[3])
\begin{eqnarray}\label{label6.1}
Z(R_2,R_1) =
\int\limits^{R_2}_{R_1}DR(t)\,\exp\Bigg(iT\int\limits^{t_2}_{t_1}\,{\rm
Tr}\,(iR\,\dot{R}\,\tau_3)\Bigg),
\end{eqnarray}
where $\dot{R} = dR/dt$ and $T = 1/2,1,3/2,\ldots$ is the colour
isospin quantum number, should be regularized by the analogy to an
axial--symmetric top. The regularized expression has been defined in Eq.(9) of Ref.[3] by
\begin{eqnarray}\label{label6.2}
Z_{\rm Reg}(R_2,R_1) =
\int\limits^{R_2}_{R_1}DR(t)\,\exp\Bigg(i\int\limits^{t_2}_{t_1}
\Big[\frac{1}{2}\,I_{\perp}\,(\Omega^2_1 + \Omega^2_2) +
\frac{1}{2}\,I_{\parallel}\,\Omega^2_3 + T\,\Omega_3\Big]\Bigg),
\end{eqnarray}
where $\Omega_a = i\,{\rm Tr}(R\,\dot{R}\,\tau_a)$ are angular
velocities of the top, $\tau_a$ are Pauli matrices $a=1,2,3$,
$I_{\perp}$ and $I_{\parallel}$ are the moments of inertia of the top
which should be taken to zero. According to the prescription of
Ref.[3] one should take first the limit $I_{\parallel} \to 0$ and
then $I_{\perp} \to 0$.

For the confirmation of the result, given in Eq.(13) of Ref.[3],  
\begin{eqnarray}\label{label6.3}
Z_{\rm Reg}(R_2,R_1) = (2T + 1)\,D^T_{TT}(R_2R^{\dagger}_1),
\end{eqnarray}
where $D^T(U)$ is a Wigner rotational matrix in the representation $T$,
the authors of Ref.[3] suggested to evaluate the evolution operator
(\ref{label6.2}) explicitly by means of the discretization of the
path integral over $R$.  The discretized form of the evolution
operator Eq.(\ref{label6.2}) is given by Eq.(14) of Ref.[3] and
reads\footnote{We are using the notations of Ref.[3]}
\begin{eqnarray}\label{label6.4}
\hspace{-0.5in}&&Z_{\rm Reg}(R_{N+1},R_0) = \lim_{\begin{array}{c} N\to
\infty\\ \delta \to 0\end{array}}{\cal N}\int\prod^{N}_{n=1}dR_n\nonumber\\ 
\hspace{-0.5in}&&\times\,\exp\Bigg[\sum^{N}_{n=0}
\Bigg(-\,i\,\frac{I_{\perp}}{2\delta}\,\Big[({\rm Tr}\,V_n\tau_1)^2 +
({\rm Tr}\,V_n\tau_2)^2\Big] - \,i\,\frac{I_{\parallel}}{2\delta}\,({\rm
Tr}\,V_n\tau_3)^2 -\,T\,({\rm Tr}\,V_n\tau_3)\Bigg)\Bigg],
\end{eqnarray}
where $R_n = R(s_n)$ with $s_n = t_1 + n\,\delta$ and $\Omega_a = i{\rm Tr}\,(R_n
R^{\dagger}_{n+1}\tau_a)/\delta$ is the discretized analogy of the
angular velocities [3] and $V_n = R_n R^{\dagger}_{n+1}$ are the
relative orientations of the top at neighbouring points. The
normalization factor ${\cal N}$ is determined by
\begin{eqnarray}\label{label6.5}
{\cal N} = \Bigg(\frac{I_{\perp}}{2\pi i
\delta}\sqrt{\frac{I_{\parallel}}{2\pi i \delta}}\,\Bigg)^{N+1}.
\end{eqnarray}
(see Eq.(19) of Ref.[3]). According to the prescription of Ref.[3] one
should take the limits $\delta \to 0$ and $I_{\parallel}, I_{\perp}
\to 0$ but keeping the ratios $I_i/\delta$, where ($ i= {\parallel},
{\perp}$), much greater than unity, $I_i/\delta \gg 1$.

Let us rewrite the exponent of the integrand of Eq.(\ref{label6.4}) in
equivalent form
\begin{eqnarray}\label{label6.6}
\hspace{-0.5in}&&Z_{\rm Reg}(R_{N+1},R_0) = \lim_{\begin{array}{c} N\to
\infty\\ \delta \to 0\end{array}}\Bigg(\frac{I_{\perp}}{2\pi i
\delta}\sqrt{\frac{I_{\parallel}}{2\pi i \delta}}\,\Bigg)^{N+1}\int\prod^{N}_{n=1}dR_n\nonumber\\ 
\hspace{-0.5in}&&\times\,\exp\Bigg[\sum^{N}_{n=0}
\Bigg(-\,i\,\frac{I_{\perp}}{2\delta}\,({\rm Tr}\,V_n\tau_a)^2 - 
\,i\,\frac{I_{\parallel}-I_{\perp}}{2\delta}\,({\rm
Tr}\,V_n\tau_3)^2 -\,T\,({\rm Tr}\,V_n\tau_3)\Bigg)\Bigg].
\end{eqnarray}
Now let us show that if $V_n$ is a rotation in the fundamental
representation of $SU(2)$, so 
\begin{eqnarray}\label{label6.7}
({\rm Tr}\,V_n\tau_a)^2
= -4 + ({\rm Tr}\,V_n)^2.
\end{eqnarray}
For this aim, first, recall that
\begin{eqnarray}\label{label6.8}
{\rm Tr}\,(V_n\tau_a) = - {\rm
Tr}\,(V^{\dagger}_n\tau_a).
\end{eqnarray}
Since $V_n$ is a rotation matrix in the fundamental representation of
$SU(2)$, it can be taken in the general standard
parameterization given by Eq.(\ref{label4.12}). By virtue of the
relation (\ref{label6.8}) we can rewrite $({\rm Tr}\,V_n\tau_a)^2$
as follows
\begin{eqnarray}\label{label6.9}
&&({\rm Tr}\,V_n\tau_a)^2 =  - {\rm Tr}\,(V_n\tau_a){\rm
Tr}\,(V^{\dagger}_n\tau_a) = - 2\, {\rm Tr}\,\Big(\Big(V_n -
\frac{1}{2}\,{\rm Tr}\,V_n\Big)\Big(V^{\dagger}_n- \frac{1}{2}\,{\rm
Tr}\,V_n\Big) \Big) =\nonumber\\ 
&&= - 2\, {\rm Tr}\,(R_n R^{\dagger}_{n+1} R_{n+1}R^{\dagger}_n) 
+ ({\rm Tr}\,V_n)^2 = - 2\,{\rm Tr}\,1+ ({\rm Tr}\,V_n)^2 
= - 4 + ({\rm Tr}\,V_n)^2.
\end{eqnarray}
By using the relation Eq.(\ref{label6.7}) we can recast the r.h.s. of
Eq.(\ref{label6.6}) into the form
\begin{eqnarray}\label{label6.10}
\hspace{-0.5in}&&Z_{\rm Reg}(R_{N+1},R_0) = \lim_{\begin{array}{c} N\to
\infty\\ \delta \to 0\end{array}} \Bigg[\Bigg(\frac{I_{\perp}}{2\pi i
\delta}\sqrt{\frac{I_{\parallel}}{2\pi i \delta}}\,\Bigg)^{N+1}\,
\exp\Bigg(i N (N+1)\,\frac{I_{\perp}}{\delta}\Bigg)\Bigg]
\nonumber\\ 
\hspace{-0.5in}&&\times\,\int\prod^{N}_{n=1}dR_n\,
\exp\Bigg[\sum^{N}_{n=0} \Bigg(-\,i\,\frac{I_{\perp}}{2\delta}\,({\rm
Tr}\,V_n)^2 -\,i\,\frac{I_{\parallel}-I_{\perp}}{2\delta}\,({\rm
Tr}\,V_n\tau_3)^2 -\,T\,({\rm Tr}\,V_n\tau_3)\Bigg)\Bigg].
\end{eqnarray}
Now let us proceed to the evaluation of the integrals over
$R_n\,(n=1,2,...,N)$. For this aim it is convenient to rewrite the
r.h.s. of Eq.(\ref{label6.10}) in the following form
\begin{eqnarray}\label{label6.11}
\hspace{-0.5in}&&Z_{\rm Reg}(R_{N+1},R_0) = \lim_{\begin{array}{c} N\to
\infty\\ \delta \to 0\end{array}} \Bigg[\Bigg(\frac{I_{\perp}}{2\pi i
\delta}\sqrt{\frac{I_{\parallel}}{2\pi i \delta}}\,\Bigg)^{N+1}\,
\exp\Bigg(i N (N+1)\,\frac{I_{\perp}}{\delta}\Bigg)\Bigg]
\nonumber\\ 
\hspace{-0.5in}&&\times\,\int\!\!\!\int\ldots \int\!\!\!\int  
dR_N\,dR_{N-1}\,\ldots\,dR_2\,dR_1\nonumber\\
\hspace{-0.5in}&&\times\,\exp\Bigg(-\,i\,\frac{I_{\perp}}{2\delta}\,
\Big[({\rm Tr}\,R_N R^{\dagger}_{N+1} )^2 + ({\rm
Tr}\,R_{N-1} R^{\dagger}_{N} )^2 + \ldots + ({\rm
Tr}\,R_2 R^{\dagger}_1 )^2 + ({\rm
Tr}\,R_1 R^{\dagger}_0 )^2\Big] \nonumber\\
\hspace{-0.5in}&&-\,i\,\frac{I_{\parallel}-I_{\perp}}{2\delta}\,\Big[({\rm
Tr}\,R_N R^{\dagger}_{N+1}\tau_3 )^2 + ({\rm
Tr}\,R_{N-1} R^{\dagger}_{N}\tau_3 )^2 + \ldots + ({\rm
Tr}\,R_2 R^{\dagger}_1\tau_3 )^2 + ({\rm
Tr}\,R_1 R^{\dagger}_0\tau_3 )^2\Big]\nonumber\\
\hspace{-0.5in}&& -T\,\Big[{\rm
Tr}\,(R_N R^{\dagger}_{N+1}\tau_3 ) + {\rm
Tr}\,(R_{N-1} R^{\dagger}_{N}\tau_3 ) + \ldots + {\rm
Tr}\,(R_2 R^{\dagger}_1\tau_3 ) + {\rm
Tr}\,(R_1 R^{\dagger}_0\tau_3 )\Big]\Bigg).
\end{eqnarray}
In the fundamental representation and the parameterization [18]
(see Appendix) we have
\begin{eqnarray}\label{label6.12}
&&{\rm Tr}\,V_n ={\rm Tr}\,(R_n R^{\dagger}_{n+1}) =\nonumber\\
&&= 2\,\cos\frac{\beta_n}{2}\,\cos\frac{\beta_{n+1}}{2}\,
\cos\Bigg(\frac{\alpha_n + \gamma_n}{2} - \frac{\alpha_{n+1} +
\gamma_{n+1}}{2}\Bigg) \nonumber\\ && +
2\,\sin\frac{\beta_n}{2}\,\sin\frac{\beta_{n+1}}{2}\,
\cos\Bigg(\frac{\alpha_n - \gamma_n}{2} - \frac{\alpha_{n+1} -
\gamma_{n+1}}{2}\Bigg)=\nonumber\\ &&= 2\,\cos\Bigg(\frac{\beta_n
-\beta_{n+1} }{2}\Bigg)\, \cos\Bigg(\frac{\alpha_n -
\alpha_{n+1}}{2}\Bigg)\,\cos\Bigg(\frac{\gamma_n -
\gamma_{n+1}}{2}\Bigg) \nonumber\\ &&-
2\,\cos\Bigg(\frac{\beta_n + \beta_{n+1}}{2}\Bigg)\,
\sin\Bigg(\frac{\alpha_n - \alpha_{n+1}}{2}\Bigg)\,
\sin\Bigg(\frac{\gamma_n - \gamma_{n+1}}{2}\Bigg),\nonumber\\
&&{\rm Tr}\,(V_n\tau_3) = {\rm
Tr}\,(R_n R^{\dagger}_{n+1}\tau_3)=\nonumber\\
&&= -
2\,i\,\cos\frac{\beta_n}{2}\,\cos\frac{\beta_{n+1}}{2}\,
\sin\Bigg(\frac{\alpha_n + \gamma_n}{2} - \frac{\alpha_{n+1} +
\gamma_{n+1}}{2}\Bigg) \nonumber\\
&&+ 2\,i\,\sin\frac{\beta_n}{2}\,\sin\frac{\beta_{n+1}}{2}\,
\sin\Bigg(\frac{\alpha_n - \gamma_n}{2} - \frac{\alpha_{n+1} -
\gamma_{n+1}}{2}\Bigg)=\nonumber\\
&&=-\, 2\,i\cos\Bigg(\frac{\beta_n
-\beta_{n+1} }{2}\Bigg)\, \cos\Bigg(\frac{\alpha_n -
\alpha_{n+1}}{2}\Bigg)\,\sin\Bigg(\frac{\gamma_n -
\gamma_{n+1}}{2}\Bigg) \nonumber\\ 
&&- 2\,i\,\cos\Bigg(\frac{\beta_n + \beta_{n+1}}{2}\Bigg)\,
\sin\Bigg(\frac{\alpha_n - \alpha_{n+1}}{2}\Bigg)\,
\cos\Bigg(\frac{\gamma_n - \gamma_{n+1}}{2}\Bigg).
\end{eqnarray}
The Haar measure $R_n$ is defined by (see Eq.(A.2)): 
\begin{eqnarray}\label{label6.13}
DR_n = \frac{1}{8\pi^2}\,\sin\,\beta_n\,d\beta_n\,d\alpha_n\,d\gamma_n.
\end{eqnarray}
Due to the assumption $I_i/\delta \gg 1$, where ($ i= {\parallel},
{\perp}$), the integrals over $R_n$ are concentrated around unit
elements. Expanding ${\rm Tr}\,(V_n)$ and ${\rm Tr}\,(V_n\tau_3)$
around unit elements we get
\begin{eqnarray}\label{label6.14}
\hspace{-0.5in}{\rm Tr}\,V_n &=& {\rm Tr}\,(R_n R^{\dagger}_{n+1}) =2
- \frac{1}{4}\,(\beta_n - \beta_{n+1})^2 - \frac{1}{4}\,(\alpha_n -
\alpha_{n+1} + \gamma_n - \gamma_{n+1})^2,\nonumber\\ \hspace{-0.5in}{\rm
Tr}\,(V_n\tau_3) &=& {\rm Tr}\,(R_n
R^{\dagger}_{n+1}\tau_3) = - \,i\,(\alpha_n -
\alpha_{n+1} + \gamma_n - \gamma_{n+1}).
\end{eqnarray}
For the subsequent integration it is convenient to make a change of variables
\begin{eqnarray}\label{label6.15}
\frac{\alpha_n + \gamma_n}{2} &\to& \gamma_n,\nonumber\\ \alpha_n -
\gamma_n &\to& \alpha_n.
\end{eqnarray}
The Jacobian of this transformation is equal to unity. After this
change of variables (\ref{label6.14}) reads
\begin{eqnarray}\label{label6.16}
\hspace{-0.5in}{\rm Tr}\,V_n &=& {\rm Tr}\,(R_n R^{\dagger}_{n+1}) =2
- \frac{1}{4}\,(\beta_n - \beta_{n+1})^2 - \,(\gamma_n -
\gamma_{n+1})^2,\nonumber\\ \hspace{-0.5in}{\rm Tr}\,(V_n\tau_3) &=&
{\rm Tr}\,(R_n R^{\dagger}_{n+1}\tau_3) = -\, 2\,\,i\,(\gamma_n -
\gamma_{n+1}).
\end{eqnarray}
Since both ${\rm Tr}\,V_n$ and ${\rm Tr}\,(V_n\tau_3)$ do not depend
on $\alpha_n$, we can integrate out $\alpha_n$. This changes only the
Haar measure as follows
\begin{eqnarray}\label{label6.17}
DR_n = \frac{1}{4\pi}\,\beta_n\,d\beta_n\,d\gamma_n.
\end{eqnarray}
The integration over $\beta_n$ and $\gamma_n$ we will carry out in the
limits $-\infty \le \beta_n \le \infty$ and $-\infty \le \gamma_n \le
\infty$.

Substituting expansions (\ref{label6.16}) in the integrand of
Eq.(\ref{label6.11}) we obtain
\begin{eqnarray}\label{label6.18}
\hspace{-0.5in}&&Z_{\rm Reg}(R_{N+1},R_0) = \lim_{\begin{array}{c}
N\to \infty\\ \delta \to 0\end{array}} \Bigg(\frac{I_{\perp}}{2\pi i
\delta}\sqrt{\frac{I_{\parallel}}{2\pi i
\delta}}\,\Bigg)^{N+1}\,\Bigg(\frac{1}{4\pi}\Bigg)^N\nonumber\\
\hspace{-0.5in}&&\times
\int\limits^{\infty}_{-\infty}d\gamma_N
\int\limits^{\infty}_{-\infty}d\beta_N\,\beta_N
\int\limits^{\infty}_{-\infty}d\gamma_{N-1}
\int\limits^{\infty}_{-\infty}d\beta_{N-1}\,\beta_{N-1}\,\ldots
\int\limits^{\infty}_{-\infty}d\gamma_2
\int\limits^{\infty}_{-\infty}d\beta_2\,\beta_2
\int\limits^{\infty}_{-\infty}d\gamma_1
\int\limits^{\infty}_{-\infty}d\beta_1\,\beta_1
\nonumber\\
\hspace{-0.5in}&&\times\,\exp\Bigg(i\,\frac{I_{\perp}}{2\delta}\,
[(\beta_{N+1} - \beta_N)^2 + (\beta_N - \beta_{N-1})^2 + \ldots
+(\beta_2 - \beta_1)^2 + (\beta_1 - \beta_0)^2] \nonumber\\
\hspace{-0.5in}&&+\,i\,\frac{I_{\parallel}}{2\delta}\,[(\gamma_{N+1} -
\gamma_N)^2 + (\gamma_N - \gamma_{N-1})^2 + \ldots +(\gamma_2 -
\gamma_1)^2 + (\gamma_1 - \gamma_0)^2]\nonumber\\
\hspace{-0.5in}&& - 2\,i\,T\,[(\gamma_{N+1} - \gamma_N) + (\gamma_N
- \gamma_{N-1}) + \ldots +(\gamma_2 - \gamma_1) + (\gamma_1 -
\gamma_0)]\Bigg)=\nonumber\\
\hspace{-0.5in}&&= e^{\textstyle -\,2\,i\,T\,(\gamma_{N+1} -
\gamma_0)}\lim_{\begin{array}{c} N\to \infty\\ \delta \to
0\end{array}} \Bigg(\frac{I_{\perp}}{2\pi i
\delta}\sqrt{\frac{I_{\parallel}}{2\pi i
\delta}}\,\Bigg)^{N+1}\,\Bigg(\frac{1}{4\pi}\Bigg)^N\nonumber\\
\hspace{-0.5in}&&\times
\int\limits^{\infty}_{-\infty}d\gamma_N
\int\limits^{\infty}_{-\infty}d\beta_N\,\beta_N
\int\limits^{\infty}_{-\infty}d\gamma_{N-1}
\int\limits^{\infty}_{-\infty}d\beta_{N-1}\,\beta_{N-1}\,\ldots
\int\limits^{\infty}_{-\infty}d\gamma_2
\int\limits^{\infty}_{-\infty}d\beta_2\,\beta_2
\int\limits^{\infty}_{-\infty}d\gamma_1
\int\limits^{\infty}_{-\infty}d\beta_1\,\beta_1
\nonumber\\
\hspace{-0.5in}&&\times\,\exp\Bigg(i\,\frac{I_{\perp}}{2\delta}\,
[(\beta_{N+1} - \beta_{N})^2 + (\beta_{N} - \beta_{N-1})^2 + \ldots
+(\beta_2 - \beta_1)^2 + (\beta_1 - \beta_0)^2] \nonumber\\
\hspace{-0.5in}&&+\,i\,\frac{I_{\parallel}}{2\delta}\,[(\gamma_{N+1} -
\gamma_{N})^2 + (\gamma_{N} - \gamma_{N-1})^2 + \ldots +(\gamma_2 -
\gamma_1)^2 + (\gamma_1 - \gamma_0)^2]\Bigg).
\end{eqnarray}
The integration over $\gamma_n$ gives
\begin{eqnarray}\label{label6.19}
\hspace{-0.3in}&&\int\limits^{\infty}_{-\infty}d\gamma_N
\int\limits^{\infty}_{-\infty}d\gamma_{N-1}
\,\ldots
\int\limits^{\infty}_{-\infty}d\gamma_2
\int\limits^{\infty}_{-\infty}d\gamma_1
\nonumber\\
\hspace{-0.3in}&&\times\,
\exp\Bigg(i\,\frac{I_{\parallel}}{2\delta}\,[(\gamma_{N+1} -
\gamma_{N})^2 + (\gamma_{N} - \gamma_{N-1})^2 + \ldots +(\gamma_2 -
\gamma_1)^2 + (\gamma_1 - \gamma_0)^2]\Bigg)=\nonumber\\
\hspace{-0.3in}&&=\sqrt{\frac{2\pi
i\delta}{I_{\parallel}}\,\frac{1}{2}}\sqrt{\frac{2\pi
i\delta}{I_{\parallel}}\,\frac{2}{3}} \ldots \sqrt{\frac{2\pi
i\delta}{I_{\parallel}}\,\frac{N-1}{N}}\sqrt{\frac{2\pi
i\delta}{I_{\parallel}}\,\frac{N}{N+1}}\,
\exp\Bigg(i\,\frac{I_{\parallel}}{2(N+1)\delta}\,(\gamma_{N+1} -
\gamma_0)^2\Bigg)=\nonumber\\
\hspace{-0.3in}&&= \Bigg(\sqrt{\frac{2\pi
i\delta}{I_{\parallel}}}\,\Bigg)^N \sqrt{\frac{1}{N+1}}\,\,
\exp\Bigg(i\,\frac{I_{\parallel}}{2(N+1)\delta}\,(\gamma_{N+1} -
\gamma_0)^2\Bigg).
\end{eqnarray}
By taking into account the normalization factor the result of the
integration over $\gamma_n$ reads
\begin{eqnarray}\label{label6.20}
\hspace{-0.5in}&&\Bigg(\sqrt{\frac{I_{\parallel}}{2\pi i
\delta}}\,\Bigg)^{N+1}\,\int\limits^{\infty}_{-\infty}d\gamma_N
\int\limits^{\infty}_{-\infty}d\gamma_{N-1}
\,\ldots
\int\limits^{\infty}_{-\infty}d\gamma_2
\int\limits^{\infty}_{-\infty}d\gamma_1
\nonumber\\
\hspace{-0.5in}&&\times\,
\exp\Bigg(i\,\frac{I_{\parallel}}{2\delta}\,[(\gamma_{N+1} -
\gamma_{N})^2 + (\gamma_{N} - \gamma_{N-1})^2 + \ldots +(\gamma_2 -
\gamma_1)^2 + (\gamma_1 - \gamma_0)^2]\Bigg)=\nonumber\\
\hspace{-0.5in}&&=\sqrt{\frac{I_{\parallel}}{2\pi i(N+1)\delta}}\,
\exp\Bigg(i\,\frac{I_{\parallel}}{2(N+1)\delta}\,(\gamma_{N+1} -
\gamma_0)^2\Bigg) = \nonumber\\
\hspace{-0.5in}&&=\sqrt{\frac{I_{\parallel}}{2\pi
i\Delta t}}\,
\exp\Bigg(i\,\frac{I_{\parallel}}{2\Delta t}\,(\gamma_{N+1} -
\gamma_0)^2\Bigg),
\end{eqnarray}
where we have replaced $(N+1)\,\delta = t_2 - t_1 = \Delta t$. The
obtained result is exact. By replacing $I_{\parallel} \to
M$, $\gamma_{N+1} \to x_b$, $\gamma_0 \to x_a$ and $\Delta t \to (t_b
- t_a)$ we arrive at the expression for the Green function, the
evolution operator, of a free particle with a mass $M$ given by
Eq.(2.51) of Ref.[6].

Thus, after the integration over $\gamma_n$ the evolution operator
$Z_{\rm Reg}(R_{N+1},R_0)$ can be written in the form
\begin{eqnarray}\label{label6.21}
\hspace{-0.5in}&&Z_{\rm Reg}(R_{N+1},R_0) =\nonumber\\
\hspace{-0.5in}&&=\sqrt{\frac{I_{\parallel}}{2\pi i\Delta t}}\,
\exp\Bigg(i\,\frac{I_{\parallel}}{2\Delta t}\,(\gamma_{N+1} -
\gamma_0)^2\Bigg)\,e^{\textstyle -\,2\,i\,T\,(\gamma_{N+1} -
\gamma_0)}\,F[I_{\perp},\beta_{N+1},\beta_0],
\end{eqnarray}
where $F[I_{\perp},\beta_{N+1},\beta_0]$ is a functional defined by
the integrals over $\beta_n$
\begin{eqnarray}\label{label6.22}
\hspace{-0.5in}&&F[I_{\perp},\beta_{N+1},\beta_0] = \nonumber\\
\hspace{-0.5in}&&=\lim_{\begin{array}{c} N\to \infty\\ \delta \to
0\end{array}} \Bigg(\frac{I_{\perp}}{2\pi i
\delta}\Bigg)^{N+1}\,\Bigg(\frac{1}{4\pi}\Bigg)^N
\int\limits^{\infty}_{-\infty}d\beta_N\,\beta_N
\int\limits^{\infty}_{-\infty}d\beta_{N-1}\,\beta_{N-1}\,\ldots
\int\limits^{\infty}_{-\infty}d\beta_2\,\beta_2
\int\limits^{\infty}_{-\infty}d\beta_1\,\beta_1 \nonumber\\
\hspace{-0.5in}&&\times\,\exp\Bigg(i\,\frac{I_{\perp}}{2\delta}\,
[(\beta_{N+1} - \beta_{N})^2 + (\beta_{N} - \beta_{N-1})^2 + \ldots
+(\beta_2 - \beta_1)^2 + (\beta_1 - \beta_0)^2]\Bigg).
\end{eqnarray}
Formally we do not need to evaluate the functional
$F[I_{\perp},\beta_{N+1},\beta_0]$ explicitly. In fact, the functional
$F[I_{\perp},\beta_{N+1},\beta_0]$ should be a regular function of
variables $I_{\perp}$, $\beta_{N+1}$ and $\beta_0$ restricted in the
limit $I_{\perp} \to 0$. Therefore, taking the limit $I_{\parallel}
\to 0$ for the evolution operator $Z_{\rm Reg}(R_{N+1},R_0)$ defined
by Eq.(\ref{label6.21}) we get 
\begin{eqnarray}\label{label6.23}
Z(R_2,R_1) = \lim_{I_{\parallel},I_{\perp} \to 0}Z_{\rm Reg}(R_{N+1},R_0) = 0.
\end{eqnarray}
This agrees with our results obtained in Sects.\,4 and 5.

Nevertheless, in spite of this very definite result let us proceed to
the explicit evaluation of the functional
$F[I_{\perp},\beta_{N+1},\beta_0]$. It is convenient to rewrite the
integrand of Eq.(\ref{label6.22}) in the equivalent form
\begin{eqnarray}\label{label6.24}
\hspace{-0.5in}&&F[I_{\perp},\beta_{N+1},\beta_0]
=\lim_{\begin{array}{c} N\to \infty\\ \delta \to 0\end{array}}
\Bigg(\frac{I_{\perp}}{2\pi i
\delta}\Bigg)\,\Bigg(\frac{-1}{\pi}\Bigg)^N\,\Bigg(\frac{1}{4\pi}\Bigg)^N
\nonumber\\
\hspace{-0.5in}&& \times\,
\frac{\partial}{\partial j_1}\frac{\partial}{\partial j_2}\ldots 
\frac{\partial}{\partial j_{N-1}}\frac{\partial}{\partial j_N}
\int\limits^{\infty}_{-\infty}d\beta_N
\int\limits^{\infty}_{-\infty}d\beta_{N-1}\ldots
\int\limits^{\infty}_{-\infty}d\beta_2
\int\limits^{\infty}_{-\infty}d\beta_1\nonumber\\
\hspace{-0.5in}&&\times\,\exp\Bigg(i\,\frac{I_{\perp}}{2\delta}\,
[(\beta_{N+1} - \beta_{N})^2 + (\beta_{N} - \beta_{N-1})^2 + \ldots
+(\beta_2 - \beta_1)^2 + (\beta_1 - \beta_0)^2\nonumber\\
\hspace{-0.5in}&& + j_N\,\beta_N +j_{N-1}\,\beta_{N-1} + \ldots + j_2\,
\beta_2 + j_1\,\beta_1 ]\Bigg)\Bigg|_{j_N = j_{N-1} = \ldots =j_2 = j_1 = 0} .
\end{eqnarray}
After $k$ integrations we get
\begin{eqnarray}\label{label6.25}
\hspace{-0.5in}&&\int\limits^{\infty}_{-\infty}d\beta_k
\int\limits^{\infty}_{-\infty}d\beta_{k-1}\ldots
\int\limits^{\infty}_{-\infty}d\beta_2
\int\limits^{\infty}_{-\infty}d\beta_1\,
\exp\Bigg(i\,\frac{I_{\perp}}{2\delta}\,
[(\beta_{k+1} - \beta_{k})^2 + (\beta_{k} - \beta_{k-1})^2\nonumber\\
\hspace{-0.5in}&& + \ldots +(\beta_2 - \beta_1)^2 + (\beta_1 -
\beta_0)^2 + j_k\,\beta_k +j_{k-1}\,\beta_{k-1} + \ldots + j_2\,
\beta_2 + j_1\,\beta_1 ]\Bigg) =\nonumber\\
\hspace{-0.5in}&&=\sqrt{\frac{2\pi
i\delta}{I_{\perp}}\,\frac{1}{2}}\sqrt{\frac{2\pi
i\delta}{I_{\perp}}\,\frac{2}{3}} \ldots \sqrt{\frac{2\pi
i\delta}{I_{\perp}}\,\frac{k-1}{k}}\sqrt{\frac{2\pi
i\delta}{I_{\perp}}\,\frac{k}{k+1}}\,
\exp\Bigg(i\,\frac{I_{\perp}}{2(k+1)\delta}\,(
\beta_0 - \beta_{k+1})^2\Bigg)\nonumber\\
\hspace{-0.5in}&& 
\times\,\exp\Bigg(i\,\frac{I_{\perp}}{\delta}\,\beta_{k+1}\,
\Bigg(\frac{k}{k+1}\,j_k +
\frac{k}{k+1}\cdot\frac{k-1}{k}\,j_{k-1}
+\frac{k}{k+1}\cdot\frac{k-1}{k}\cdot\frac{k-2}{k-1}\,j_{k-2}\nonumber\\
\hspace{-0.5in}&& + \ldots +
\frac{k}{k+1}\cdot\frac{k-1}{k}\cdot\frac{k-2}{k-1}\cdots
\frac{2}{3}\,j_2 +
\frac{k}{k+1}\cdot\frac{k-1}{k}\cdot\frac{k-2}{k-1}\cdots
\frac{2}{3}\cdot\frac{1}{2}\,j_1\Bigg)\Bigg)\nonumber\\
\hspace{-0.5in}&&\times\,\exp\Bigg(i\,\frac{I_{\perp}}{\delta}\,
\Bigg[-\frac{1}{2}\cdot\frac{1}{2}\,\Big(j_1 - \beta_0\Big)^2 +
\frac{1}{2}\cdot\frac{1}{2}\,\beta^2_0 -
\frac{1}{2}\cdot\frac{2}{3}\,\Big(j_2 + \frac{1}{2}\,j_1 -
\frac{1}{2}\,\beta_0\Big)^2 +
\frac{1}{2}\cdot\frac{2}{3}\cdot\frac{1}{2^2}\,\beta^2_0\nonumber\\
\hspace{-0.5in}&& -\frac{1}{2}\cdot\frac{3}{4}\,\Big(j_3 +
\frac{2}{3}\, j_2 + \frac{2}{3}\cdot\frac{1}{2}\,j_1 -
\frac{1}{3}\,\beta_0\Big)^2 +
\frac{1}{2}\cdot\frac{3}{4}\cdot\frac{1}{3^2}\,\beta^2_0 -
\frac{1}{2}\cdot\frac{4}{5}\,\Big(j_4 + \frac{3}{4}\,j_3 +
\frac{3}{4}\cdot\frac{2}{3}\, j_2\nonumber\\
\hspace{-0.5in}&& + \frac{3}{4}\cdot\frac{2}{3}\cdot\frac{1}{2}\,j_1 -
\frac{1}{4}\,\beta_0\Big)^2 +
\frac{1}{2}\cdot\frac{4}{5}\cdot\frac{1}{4^2}\,\beta^2_0 -
\frac{1}{2}\cdot\frac{5}{6}\,\Big(j_5 + \frac{4}{5}\,j_4
+\frac{4}{5}\cdot \frac{3}{4}\,j_3 + \frac{4}{5}\cdot
\frac{3}{4}\cdot\frac{2}{3}\, j_2\nonumber\\
\hspace{-0.5in}&& +
\frac{4}{5}\cdot\frac{3}{4}\cdot\frac{2}{3}\cdot\frac{1}{2}\,j_1 -
\frac{1}{5}\,\beta_0\Big)^2 +
\frac{1}{2}\cdot\frac{5}{6}\cdot\frac{1}{5^2}\,\beta^2_0 -
\ldots - \frac{1}{2}\cdot\frac{k}{k+1}\,\Big(j_k +
\frac{k-1}{k}\,j_{k-1} \nonumber\\
\hspace{-0.5in}&& + \frac{k-1}{k}\cdot\frac{k-2}{k-1}\,j_{k-2} +
\ldots + \frac{k-1}{k}\cdot\frac{k-2}{k-1}\cdots \frac{2}{3}\,j_2 + \frac{k-1}{k}\cdot\frac{k-2}{k-1}\cdots
\frac{2}{3}\cdot\frac{1}{2}\,j_1 - \frac{1}{k}\,\beta_0\Big)^2 \nonumber\\
\hspace{-0.5in}&& +
\frac{1}{2}\cdot\frac{k}{k+1}\cdot\frac{1}{k^2}\,\beta^2_0\Bigg]\Bigg).
\end{eqnarray}
By performing $N$ integrations we obtain
\begin{eqnarray}\label{label6.26}
\hspace{-0.3in}&&\int\limits^{\infty}_{-\infty}d\beta_N
\int\limits^{\infty}_{-\infty}d\beta_{N-1}\ldots
\int\limits^{\infty}_{-\infty}d\beta_2
\int\limits^{\infty}_{-\infty}d\beta_1\,
\exp\Bigg(i\,\frac{I_{\perp}}{2\delta}\,
[(\beta_{N+1} - \beta_{N})^2 + (\beta_{N} - \beta_{N-1})^2\nonumber\\
\hspace{-0.3in}&& + \ldots +(\beta_2 - \beta_1)^2 + (\beta_1 -
\beta_0)^2 + j_N\,\beta_N +j_{N-1}\,\beta_{N-1} + \ldots + j_2\,
\beta_2 + j_1\,\beta_1 ]\Bigg) =\nonumber\\
\hspace{-0.3in}&&=\Bigg(\sqrt{\frac{2\pi
i\delta}{I_{\perp}}}\,\Bigg)^{N+1}\,\sqrt{\frac{I_{\perp}}{2\pi
i\Delta t}}\,
\exp\Bigg(i\,\frac{I_{\perp}}{2\Delta t}\,(
\beta_{N+1} - \beta_0)^2\Bigg)\nonumber\\
\hspace{-0.3in}&& 
\times\,\exp\Bigg(i\,\frac{I_{\perp}}{\delta}\,\beta_{N+1}\,
\Bigg(\frac{N}{N+1}\,j_N +
\frac{N}{N+1}\cdot\frac{N-1}{N}\,j_{N-1}
+\frac{N}{N+1}\cdot\frac{N-1}{N}\cdot\frac{N-2}{N-1}\,j_{N-2}\nonumber\\
\hspace{-0.5in}&& + \ldots +
\frac{N}{N+1}\cdot\frac{N-1}{N}\cdot\frac{N-2}{N-1}\cdots
\frac{2}{3}\,j_2 +
\frac{N}{N+1}\cdot\frac{N-1}{N}\cdot\frac{N-2}{N-1}\cdots
\frac{2}{3}\cdot\frac{1}{2}\,j_1\Bigg)\Bigg)\nonumber\\
\hspace{-0.3in}&&\times\,\exp\Bigg(i\,\frac{I_{\perp}}{\delta}\,
\Bigg[-\frac{1}{2}\cdot\frac{1}{2}\,\Big(j_1 - \beta_0\Big)^2 +
\frac{1}{2}\cdot\frac{1}{2}\,\beta^2_0 -
\frac{1}{2}\cdot\frac{2}{3}\,\Big(j_2 + \frac{1}{2}\,j_1 -
\frac{1}{2}\,\beta_0\Big)^2 +
\frac{1}{2}\cdot\frac{2}{3}\cdot\frac{1}{2^2}\,\beta^2_0\nonumber\\
\hspace{-0.3in}&& -\frac{1}{2}\cdot\frac{3}{4}\,\Big(j_3 +
\frac{2}{3}\, j_2 + \frac{2}{3}\cdot\frac{1}{2}\,j_1 -
\frac{1}{3}\,\beta_0\Big)^2 +
\frac{1}{2}\cdot\frac{3}{4}\cdot\frac{1}{3^2}\,\beta^2_0 -
\frac{1}{2}\cdot\frac{4}{5}\,\Big(j_4 + \frac{3}{4}\,j_3 +
\frac{3}{4}\cdot\frac{2}{3}\, j_2\nonumber\\
\hspace{-0.3in}&& + \frac{3}{4}\cdot\frac{2}{3}\cdot\frac{1}{2}\,j_1 -
\frac{1}{4}\,\beta_0\Big)^2 +
\frac{1}{2}\cdot\frac{4}{5}\cdot\frac{1}{4^2}\,\beta^2_0 -
\frac{1}{2}\cdot\frac{5}{6}\,\Big(j_5 + \frac{4}{5}\,j_4 +
\frac{4}{5}\cdot \frac{3}{4}\,j_3 + \frac{4}{5}\cdot
\frac{3}{4}\cdot\frac{2}{3}\, j_2\nonumber\\
\hspace{-0.3in}&& +
\frac{4}{5}\cdot\frac{3}{4}\cdot\frac{2}{3}\cdot\frac{1}{2}\,j_1 -
\frac{1}{5}\,\beta_0\Big)^2 +
\frac{1}{2}\cdot\frac{5}{6}\cdot\frac{1}{5^2}\,\beta^2_0 -
\ldots - \frac{1}{2}\cdot\frac{N}{N+1}\,\Big(j_N +
\frac{N-1}{N}\,j_{N-1} \nonumber\\
\hspace{-0.3in}&& + \frac{N-1}{N}\cdot\frac{N-2}{N-1}\,j_{N-2} +
\ldots + \frac{N-1}{N}\cdot\frac{N-2}{N-1}\cdots \frac{2}{3}\,j_2 +
\frac{N-1}{N}\cdot\frac{N-2}{N-1}\cdots
\frac{2}{3}\cdot\frac{1}{2}\,j_1
\nonumber\\
\hspace{-0.3in}&& - \frac{1}{N}\,\beta_0\Big)^2  +
\frac{1}{2}\cdot\frac{N}{N+1}\cdot\frac{1}{N^2}\,\beta^2_0\Bigg]\Bigg),
\end{eqnarray}
where we have replaced $(N+1)\,\delta = t_2 - t_1 = \Delta t$.

Now we can evaluate the derivatives with respect to $j_1, j_2, \ldots
,j_{N-1},j_N$. Due to the constraint $I_{\perp}/\delta \gg 1$ we can
keep only the leading order contributions in powers of
$I_{\perp}/\delta \gg 1$. The result reads
\begin{eqnarray}\label{label6.27}
\hspace{-0.5in}&&\frac{\partial}{\partial j_1}\frac{\partial}{\partial
j_2}\ldots \frac{\partial}{\partial j_{N-1}}\frac{\partial}{\partial
j_N} \int\limits^{\infty}_{-\infty}d\beta_N
\int\limits^{\infty}_{-\infty}d\beta_{N-1}\ldots
\int\limits^{\infty}_{-\infty}d\beta_2
\int\limits^{\infty}_{-\infty}d\beta_1\nonumber\\
\hspace{-0.5in}&&\times\,\exp\Bigg(i\,\frac{I_{\perp}}{2\delta}\,
[(\beta_{N+1} - \beta_{N})^2 + (\beta_{N} - \beta_{N-1})^2 + \ldots
+(\beta_2 - \beta_1)^2 + (\beta_1 - \beta_0)^2\nonumber\\
\hspace{-0.5in}&& + j_N\,\beta_N +j_{N-1}\,\beta_{N-1} + \ldots +
j_2\, \beta_2 + j_1\,\beta_1 ]\Bigg)\Bigg|_{j_N = j_{N-1} = \ldots
=j_2 = j_1 = 0} =\nonumber\\
\hspace{-0.5in}&&= \Bigg(\sqrt{\frac{2\pi
i\delta}{I_{\perp}}}\,\Bigg)^{N+1}\,\sqrt{\frac{I_{\perp}}{2\pi
i\Delta t}}\, \exp\Bigg(i\,\frac{I_{\perp}}{2\Delta t}\,( \beta_{N+1}
- \beta_0)^2\Bigg)\,\Bigg(\frac{iI_{\perp}}{\delta}\Bigg)^N\,\nonumber\\
\hspace{-0.5in}&&\Bigg(\beta_{N+1}\,\frac{1}{N+1}+\beta_0\,\Bigg[\frac{1}{1\cdot 2} + \frac{1}{2\cdot 3} +
\frac{1}{3\cdot 4} + \ldots +  \frac{1}{N(N+1)}\Bigg]\Bigg)\,\nonumber\\
\hspace{-0.5in}&&\times\,\Bigg(\beta_{N+1}\,\frac{2}{N+1} + \beta_0\,\Bigg[\frac{2}{2\cdot 3} +
\frac{2}{3\cdot 4} + \ldots +  \frac{2}{N(N+1)}\Bigg]\Bigg)\,\nonumber\\
\hspace{-0.5in}&&\Bigg(\beta_{N+1}\,\frac{3}{N+1} +
\beta_0\,\Bigg[\frac{3}{3\cdot 4} + \ldots +
\frac{3}{N(N+1)}\Bigg]\Bigg)\,\nonumber\\
\hspace{-0.5in}&&\times\,\Bigg(\beta_{N+1}\,\frac{4}{N+1} +
\beta_0\, \Bigg[\frac{4}{4\cdot5 } +\ldots +
\frac{4}{N(N+1)}\Bigg]\Bigg)\ldots \Bigg(\beta_{N+1}\,\frac{N}{N+1} +
\beta_0\,\frac{N}{N(N+1)}\Bigg)=\nonumber\\
\hspace{-0.5in}&&= \Bigg(\sqrt{\frac{2\pi
i\delta}{I_{\perp}}}\,\Bigg)^{N+1}\,\sqrt{\frac{I_{\perp}}{2\pi
i\Delta t}}\, \exp\Bigg(i\,\frac{I_{\perp}}{2\Delta t}\,( \beta_{N+1}
- \beta_0)^2\Bigg)\,\Bigg(\frac{iI_{\perp}}{\delta}\Bigg)^N\,\nonumber\\
\hspace{-0.5in}&&\times\,\Bigg(\beta_{N+1}\,\frac{1}{N+1} +
\beta_0\,\frac{N}{N+1}\Bigg)\,\Bigg(\beta_{N+1}\,\frac{2}{N+1} +
\beta_0\,\frac{N-1}{N+1}\Bigg)\nonumber\\
\hspace{-0.5in}&&\times\,\Bigg(\beta_{N+1}\,\frac{3}{N+1} +
\beta_0\,\frac{N-2}{N+1}\Bigg)\,\times\,\ldots\times\,
\Bigg(\beta_{N+1}\,\frac{N}{N+1}
+ \beta_0\,\frac{1}{N+1}\Bigg)=\nonumber\\
\hspace{-0.5in}&&= \Bigg(\sqrt{\frac{2\pi
i\delta}{I_{\perp}}}\,\Bigg)^{N+1}\,\sqrt{\frac{I_{\perp}}{2\pi
i\Delta t}}\, \exp\Bigg(i\,\frac{I_{\perp}}{2\Delta t}\,( \beta_{N+1}
-  \beta_0)^2\Bigg)\,\Bigg(\frac{iI_{\perp}}{\delta}\Bigg)^N\,\nonumber\\
\hspace{-0.5in}&&\times\,\prod^{N}_{k=1}\Bigg(\beta_{N+1}\,\frac{k}{N+1} +
\beta_0\,\frac{N+1-k}{N+1}\Bigg).
\end{eqnarray}
Substituting Eq.(\ref{label6.27}) in Eq.(\ref{label6.24}) we obtain
the functional $F[I_{\perp},\beta_{N+1},\beta_0]$: 
\begin{eqnarray}\label{label6.28}
\hspace{-0.5in}&&F[I_{\perp},\beta_{N+1},\beta_0]
= \sqrt{\frac{I_{\perp}}{2\pi i\Delta t}}\,
\exp\Bigg(i\,\frac{I_{\perp}}{2\Delta t}\,( \beta_{N+1} -
\beta_0)^2\Bigg)\nonumber\\
\hspace{-0.5in}&&\times\,\lim_{\begin{array}{c} N\to \infty\\ \delta \to
0\end{array}}\Bigg(\frac{1}{2\pi}\Bigg)^N \Bigg(\sqrt{\frac{I_{\perp}}{2\pi i
\delta}}\,\Bigg)^{N+1}\,\prod^{N}_{k=1}\Bigg(\beta_{N+1}\,\frac{k}{N+1} +
\beta_0\,\frac{N+1-k}{N+1}\Bigg).
\end{eqnarray}
In order to understand the behaviour of the functional
$F[I_{\perp},\beta_{N+1},\beta_0]$ in the limit $N \to \infty$ we
suggest to evaluate the product
\begin{eqnarray}\label{label6.29}
{\Pi}[\beta_{N+1},\beta_0] = \prod^{N}_{k=1}\Bigg(\beta_{N+1}\,\frac{k}{N+1} +
\beta_0\,\frac{N+1-k}{N+1}\Bigg)
\end{eqnarray}
at $N \gg 1$ by using the $\zeta$--regularization. In the $\zeta$--regularization the evaluation of ${\Pi}[\beta_{N+1},\beta_0]$ runs the following way
\begin{eqnarray}\label{label6.30}
&&{\ell n}\,{\Pi}[\beta_{N+1},\beta_0] = \sum^{N}_{k=1}{\ell
n}\,\Bigg(\beta_{N+1}\,\frac{k}{N+1} +
\beta_0\,\frac{N+1-k}{N+1}\Bigg) = \nonumber\\
&&=\sum^{N}_{k=1}(-1)\,\frac{d}{ds}\Bigg(\beta_{N+1}\,\frac{k}{N+1} +
\beta_0\,\frac{N+1-k}{N+1}\Bigg)^{-s}\Bigg|_{s=0}=\nonumber\\
&&=-\frac{d}{ds}\sum^{N}_{k=1}\int\limits^{\infty}_0\frac{dz}{\Gamma(s)}\,
\exp\Bigg[-\Bigg(\beta_{N+1}\,\frac{k}{N+1} +
\beta_0\,\frac{N+1-k}{N+1}\Bigg)\,z\Bigg]\,z^{s-1}\Bigg|_{s=0}=\nonumber\\
&&=-\frac{d}{ds} \int\limits^{\infty}_0\frac{dz}{\Gamma(s)}\,
\left[\frac{\displaystyle e^{\textstyle -\beta_0\,z} - e^{\textstyle -
\beta_{N+1}\,z}}{\displaystyle 1 - \exp \Big( -\frac{\beta_{N+1} -
\beta_0}{N+1}\,z\Big)}- e^{\textstyle
-\beta_0\,z}\right]\,z^{s-1}\Bigg|_{s=0} =\nonumber\\ &&=-
(N+1)\,\frac{d}{ds} \int\limits^{\infty}_0\frac{dz}{\Gamma(s)}\,
\frac{\displaystyle e^{\textstyle -\beta_0\,z} - e^{\textstyle -
\beta_{N+1}\,z}}{\beta_{N+1} - \beta_0}\,z^{s-2}\Bigg|_{s=0}
=\nonumber\\ 
&&=- (N+1)\,\frac{d}{ds}\,\left[ \frac{1}{s-1}\,
\frac{\displaystyle \beta^{1-s}_0 - \beta^{1-s}_{N+1}}{\beta_{N+1} -
\beta_0}\right]_{s=0} =\nonumber\\
&&= - (N+1)\,\Bigg[1 - \frac{\beta_{N+1}\,{\ell
n}\,\beta_{N+1} - \beta_0\,{\ell n}\,\beta_0}{\beta_{N+1} - \beta_0}\Bigg]
=\nonumber\\
 &&=- (N+1)\,\frac{\displaystyle \beta_{N+1}\,{\ell
n}\,\frac{e}{\beta_{N+1}} - \beta_0\,{\ell
n}\,\frac{e}{\beta_0}}{\beta_{N+1} - \beta_0}.
\end{eqnarray}
Thus, the function ${\Pi}[\beta_{N+1},\beta_0]$ is defined by 
\begin{eqnarray}\label{label6.31}
{\Pi}[\beta_{N+1},\beta_0] = \exp\left( - (N+1)\,\frac{\displaystyle
\beta_{N+1}\,{\ell n}\,\frac{e}{\beta_{N+1}} - \beta_0\,{\ell
n}\,\frac{e}{\beta_0}}{\beta_{N+1} - \beta_0}\,\right),
\end{eqnarray}
where $e = 2.71828\ldots$. Due to the constraint $I_{\perp}/\delta
\gg 1$ the Euler angles $\beta_{N+1}$ and $\beta_0$ are less than
unity and the ratio in Eq.(\ref{label6.31}) is always positive
\begin{eqnarray}\label{label6.32}
\frac{\displaystyle
\beta_{N+1}\,{\ell n}\,\frac{e}{\beta_{N+1}} - \beta_0\,{\ell
n}\,\frac{e}{\beta_0}}{\beta_{N+1} - \beta_0} > 0.
\end{eqnarray}
The functional $F[I_{\perp},\beta_{N+1},\beta_0]$ is then defined by 
\begin{eqnarray}\label{label6.33}
\hspace{-0.5in}&&F[I_{\perp},\beta_{N+1},\beta_0]
= \sqrt{\frac{I_{\perp}}{2\pi i\Delta t}}\,
\exp\Bigg(i\,\frac{I_{\perp}}{2\Delta t}\,( \beta_{N+1} -
\beta_0)^2\Bigg)\nonumber\\
\hspace{-0.5in}&&\times\,\lim_{\begin{array}{c} N\to \infty\\ \delta \to
0\end{array}}\Bigg(\frac{1}{2\pi}\Bigg)^N \Bigg(\sqrt{\frac{I_{\perp}}{2\pi i
\delta}}\,\Bigg)^{N+1}\,\exp\left( - (N+1)\,\frac{\displaystyle
\beta_{N+1}\,{\ell n}\,\frac{e}{\beta_{N+1}} - \beta_0\,{\ell
n}\,\frac{e}{\beta_0}}{\beta_{N+1} - \beta_0}\,\right).
\end{eqnarray}
Thus $F[I_{\perp},\beta_{N+1},\beta_0]$ vanishes in the limit $N \to
\infty$. This result retains itself even if we change the
normalization factor of the evolution operator
\begin{eqnarray}\label{label6.34}
{\cal N} = \Bigg(\frac{I_{\perp}}{2\pi i
\delta}\sqrt{\frac{I_{\parallel}}{2\pi i \delta}}\,\Bigg)^{N+1} \to
(2\pi)^N\,\Bigg(\sqrt{\frac{I_{\parallel}}{I_{\perp}}}\,\Bigg)^{N+1} .
\end{eqnarray}
The renormalized functional $F[I_{\perp},\beta_{N+1},\beta_0]$,
defined by
\begin{eqnarray}\label{label6.35}
F[I_{\perp},\beta_{N+1},\beta_0]
&=& \sqrt{\frac{I_{\perp}}{2\pi i\Delta t}}\,
\exp\Bigg(i\,\frac{I_{\perp}}{2\Delta t}\,( \beta_{N+1} -
\beta_0)^2\Bigg) \nonumber\\
&\times&\lim_{N\to \infty}\exp\left( - (N+1)\,\frac{\displaystyle
\beta_{N+1}\,{\ell n}\,\frac{e}{\beta_{N+1}} - \beta_0\,{\ell
n}\,\frac{e}{\beta_0}}{\beta_{N+1} - \beta_0}\,\right),
\end{eqnarray}
vanishes in the limit $N\to \infty$ since the Euler angles
$\beta_{N+1}$ and $\beta_0$ are small compared with unity due to the
constraint $I_{\perp}/\delta \gg 1$ [3]. The vanishing of the
functional $F[I_{\perp},\beta_{N+1},\beta_0]$ in the limit $N \to
\infty$ agrees with our results obtained in Sects.\,4 and 5.

Substituting Eq.(\ref{label6.35}) in Eq.(\ref{label6.21}) we obtain
\begin{eqnarray}\label{label6.36}
Z_{\rm Reg}(R_{N+1},R_0) = 0.
\end{eqnarray}
This leads to the vanishing of the evolution operator $Z(R_2,R_1)$
given by Eq.(\ref{label6.1}) or Eq.(8) of Ref.[3], $Z(R_2,R_1) = 0$.

Thus, the evolution operator $Z(R_2,R_1)$ suggested 
in Ref.[3] for the description of Wilson loops in terms of
path integrals over gauge degrees of freedom is equal to zero
identically.  This agrees with our results obtained in Sects.\,4 and
5. As we have shown the vanishing of $Z(R_2,R_1)$ does not depend on
the specific regularization and discretization of the path
integral. In fact, this is an intrinsic property of the path integral
given by Eq.(\ref{label6.1}) that becomes obvious if the evaluation is
carried out correctly.

\section{Evolution
 operator $Z(R_2,R_1)$ and shift of energy levels of 
an axial--symmetric top} 
\setcounter{equation}{0}

In this Section we criticize the analysis of the evolution operator
$Z(R_2,R_1)$ carried out by Diakonov and Petrov via {\it the canonical
quantization of the axial--symmetric top} (see Eq.(12) of
Ref.[3]). Below we use the notations of Ref.[3].

The parallel transport operator 
\begin{eqnarray}\label{label7.1}
W_{\alpha\beta}(t_2,t_1) =
\Bigg[P\exp\Bigg(i\int\limits^{x(t_2)}_{x(t_1)}
A^a_{\mu}(x)\,T^a\,dx_{\mu}\Bigg)\Bigg]_{\alpha\beta}
=\Bigg[P\exp\Bigg(i\int\limits^{t_2}_{t_1}
A(t)\,dt\Bigg)\Bigg]_{\alpha\beta},
\end{eqnarray}
where $A(t) = A^a_{\mu}(x)\,T^a\,dx_{\mu}/dt$ is a tangent component of
the Yang--Mills field and $T^a\,(a=1,2,3)$ are the generators of
$SU(2)$ group in the representation $T$, has been reduced to the form
\begin{eqnarray}\label{label7.2}
W_{\alpha\beta}(t_2,t_1) = D^T_{\alpha\beta}(U(t_2)U^{\dagger}(t_1)),
\end{eqnarray}
(see Eq.(5) of Ref.[3]) due to the statement [3]: {\it The potential
$A(t)$ along a given curve can be always written as a ``pure gauge''}
\begin{eqnarray}\label{label7.3}
A_{\alpha\beta}(t) =i\,
D^T_{\alpha\gamma}(U(t))\,\frac{d}{dt}D^T_{\gamma\beta}(U^{\dagger}(t))
\end{eqnarray}
(see Eq.(4) of Ref.[3]).

By using  the parallel transport operator Eq.(\ref{label7.2}) the
Wilson loop $W_T(C)$ in the representation $T$ has been defined by
\begin{eqnarray}\label{label7.4}
W_T(C) = \sum_{\alpha}W_{\alpha\alpha}(t_2,t_1) =\sum_{\alpha}
D^T_{\alpha\alpha}(U(t_2)U^{\dagger}(t_1)).
\end{eqnarray}
(see Eq.(25) of Ref.[3]).  In terms of the evolution operator
$Z(R_2,R_1)$ given by Eq.(\ref{label6.1}) (see Eq.(8) of Ref.[3]) the
parallel transport operator $W_{\alpha\beta}(t_2,t_1)$ has been
recast into the form
\begin{eqnarray}\label{label7.5}
W^{\rm DP}_{\alpha\beta}(t_2,t_1) =\int\!\!\!\int dR_1\,dR_2\sum_{T', m}(2T' +
1)\, D^{T'}_{\alpha
m}(U(t_2)R^{\dagger}_2)D^{T'}_{m\beta}(R_1U^{\dagger}(t_1))\,Z(R_2,R_1),
\end{eqnarray}
where the index ${\rm DP}$ means that the parallel transport operator
is taken in the Diakonov--Petrov (DP) representation. The Wilson loop
$W^{\rm DP}_T(C)$ in the DP--representation reads
\begin{eqnarray}\label{label7.6}
W^{\rm DP}_T(C) = \int\!\!\!\int dR_1\,dR_2\sum_{T', m, \alpha}(2T' +
1)\, D^{T'}_{\alpha
m}(U(t_2)R^{\dagger}_2)D^{T'}_{m\alpha}(R_1U^{\dagger}(t_1))\,Z(R_2,R_1).
\end{eqnarray}
Of course, if the DP--representation were correct we should get $W^{\rm
DP}_T(C) = W_T(C)$, where $ W_T(C)$ is determined by
Eq.(\ref{label7.4}).

 The regularized evolution operator $Z_{\rm Reg}(R_2,R_1)$ given by
 Eq.(\ref{label6.3}) (see also Eq.(9) of Ref.[3]) can be represented
 in the form of {\it a sum over possible intermediate states},
 i.e. eigenfunctions of the axial--symmetric top
\begin{eqnarray}\label{label7.7}
Z_{\rm Reg}(R_2,R_1) =\sum_{J, m, k}(2J + 1)\, D^J_{m
k}(R_2)D^J_{k m}(R^{\dagger}_1)\,e^{\textstyle -i(t_2-t_1)\,E_{J m}},
\end{eqnarray}
(see Eq.(12) of Ref.[3]), where $E_{J m}$ are the eigenvalues of the
Hamiltonian of the axial--symmetric top 
\begin{eqnarray}\label{label7.8}
E_{J m} = \frac{J(J+1) - m^2}{2 I_{\perp}} + \frac{(m - T)^2}{2
I_{\parallel}}
\end{eqnarray}
(see Eq.(11) of Ref.[3]).

As has been stated in Ref.[3]: {\it If we now take to zero
$I_{{\perp},{\parallel}} \to 0$ (first $I_{\parallel}$, then
$I_{\perp}$) we see that in the sum (12) only the lowest energy
intermediate state survives with $m=J=T$. The resulting phase factor
from the lowest energy state can be absorbed in the normalization
factor in eq.(9) since that corresponds to a shift in the energy
scale.}

The statement concerning the possibility to absorb the fluctuating
factor $\exp[-i(t_2-t_1)\,T/2 I_{\perp}]$ in the normalization 
of the path integral representing the evolution operator is the main
one allowing the r.h.s. of Eqs.(\ref{label7.5}) and (\ref{label7.6})
to escape from the vanishing in the limit $I_{\perp}\to 0$. 

In reality such a removal of the fluctuating factor is prohibited
since this leads to the change of the starting symmetry of
the system from $SU(2)$ to $U(2)$. In order to make this more 
transparent we suggest to insert $Z_{\rm Reg}(R_2,R_1)$ 
of Eq.(\ref{label7.7}) into Eq.(\ref{label7.6}) and to
express the Wilson loop $W^{\rm DP}_T(C)$ in terms of {\it a sum
over possible intermediate states}, the eigenfunctions of the
axial--symmetric top. The main idea of this substitution is the
following: as the Wilson loop is a physical quantity which can be
measured, all irrelevant normalization factors should be canceled for
the evaluation of it. Therefore, if the oscillating factor
$\exp[-\,i\,(t_2 - t_1)\,T/2I_{\perp}]$ can be really removed by a
renormalization of something, the Wilson loop should not depend on
this factor.

Substituting Eq.(\ref{label7.7}) in Eq.(\ref{label7.6}) and
integrating over $R_1$ and $R_2$ we obtain the following expansion
for the parallel transport operator in the DP--representation
\begin{eqnarray}\label{label7.9}
W^{\rm DP}_{\alpha\beta}(t_2,t_1)
=\sum_{T'}\sum^{T'}_{m=-T'}
D^{T'}_{\alpha\beta}(U(t_2)U^{\dagger}(t_1))\,e^{\textstyle
-i(t_2-t_1)\,E_{T' m}}.
\end{eqnarray}
Setting $\alpha = \beta$ and summing over $\alpha$ we get the
DP--representation for Wilson loops
\begin{eqnarray}\label{label7.10}
W^{\rm DP}_T(C) = \sum_{\alpha}W^{\rm DP}_{\alpha\alpha}(t_2,t_1)
=\sum_{T'}\sum^{T'}_{m=-T'} D^{T'}_{\alpha\alpha}(U(t_2) 
U^{\dagger}(t_1))\,e^{\textstyle - i(t_2-t_1)\,E_{T'  m}}.
\end{eqnarray}
Due  to the definition (\ref{label7.4}) the r.h.s. of
Eq.(\ref{label7.10}) can be rewritten in the form 
\begin{eqnarray}\label{label7.11}
W^{\rm DP}_T(C) = \sum_{\alpha}W^{\rm DP}_{\alpha\alpha}(t_2,t_1)
=\sum_{T'}\sum^{T'}_{m=-T'}W_{T'}(C)\,e^{\textstyle -i(t_2-t_1)\,E_{T'
m}}, 
\end{eqnarray}
where $W_{T'}(C)$ is the Wilson loop in the $T'$ representation
determined by Eq.(\ref{label7.4}).  The relation
(\ref{label7.11}) agrees to some extent with our expansion given
by Eq.(\ref{label7.1}).

Following Ref.[3] and taking the limit $I_{\parallel} \to 0$ we obtain $m
= T$. This reduces the r.h.s. of Eq.(\ref{label7.11}) to the form
\begin{eqnarray}\label{label7.12}
W^{\rm DP}_T(C) = \sum_{T'}W_{T'}(C)\,\exp\Bigg[ -i(t_2-t_1)\,\frac{T'(T' + 1)
- T^2}{2 I_{\perp}}\Bigg].
\end{eqnarray}
Now according to the prescription of Ref.[3] we should take the limit
$I_{\perp} \to 0$. Following again Ref.[3] and setting $T' = T$ we arrive at
the relation
\begin{eqnarray}\label{label7.13}
W^{\rm DP}_T(C) &=& W_T(C)\,\exp\Bigg[ -i(t_2-t_1)\,\frac{T}{2
I_{\perp}}\Bigg]= \nonumber\\
&=&\sum_{\alpha}
D^T_{\alpha\alpha}(U(t_2)U^{\dagger}(t_1))\,\exp\Bigg[
-i(t_2-t_1)\,\frac{T}{2 I_{\perp}}\Bigg].
\end{eqnarray}
Thus, Wilson loops in the DP--representation differ from original
 Wilson loops by the oscillating factor $\exp[ i(t_2-t_1)\,T/2
I_{\perp}]$. The only possibility to remove the oscillating factor
$\exp[ -i(t_2-t_1)\,T/2 I_{\perp}]$ is to absorb this phase
factors in the matrices $U(t_2)$ and $U^{\dagger}(t_1)$ which describe
the degrees of freedom of the gauge potential $A(t)$ via the relation
(\ref{label7.2}). This yields the changes
\begin{eqnarray}\label{label7.14}
U(t_2) &\to& \bar{U}(t_2) = U(t_2)\,e^{\textstyle i\,t_2\,T/2
I_{\perp}},\nonumber\\ U^{\dagger}(t_1) &\to& \bar{U}^{\dagger}(t_1) =
U^{\dagger}(t_1)\,e^{\textstyle -\,i\,t_1\,T/2 I_{\perp}}.
\end{eqnarray}
However, the matrices $\bar{U}(t_2)$ and $\bar{U}^{\dagger}(t_1)$ are
now the elements of the $U(2)$ group but not of $SU(2)$. Thus, the
shift of the energy level of the ground state of the axial--symmetric
top suggested by Diakonov and Petrov in order to remove the
oscillating factor changes crucially the starting symmetry of the
theory from $SU(2)$ to $U(2)$.  Since the former is
not allowed the oscillating factor $\exp[ -i(t_2-t_1)\,T/2 I_{\perp}]$
cannot be removed. As a result in the limit $I_{\perp} \to 0$ we
obtain
\begin{eqnarray}\label{label7.15}
W^{\rm DP}_T(C) = 0.
\end{eqnarray}
The vanishing of Wilson loops in the DP--representation agrees with
our results obtained in Sects.\,4, 5 and 6 and 
confirms our claim that this path integral representation of
Wilson loops is incorrect.

\section{The non--Abelian Stokes theorem}
\setcounter{equation}{0}

The derivation of the  area--law falloff promoted great interests
in  the non--Abelian Stokes theorem expressing the
exponent of Wilson loops in terms of a  surface integral over the
 2--dimensional surface $S$ with the  boundary $C = \partial S$ [19]
\begin{eqnarray}\label{label8.1}
\hspace{-0.2in}{\rm tr}\,{\cal P}_Ce^{\textstyle i\,g\,\oint_C d
x_{\mu}\,A_{\mu}(x)} = {\rm tr}\,{\cal P}_S\,
e^{\textstyle  i\,g\,\frac{1}{2}
\int\!\!\!\int\limits_S 
d\sigma_{\mu\nu}(y)\,U(C_{xy})\,G_{\mu\nu}(y)\,U(C_{yx})},
\end{eqnarray}
where ${\cal P}_S$ is the surface ordering operator [19],
$d\sigma^{\mu\nu}(y)$ is a 2--dimensional surface element in
4--dimensional space--time, $x$ is a current point on the contour $C$,
i.e. $x \in C$, $y$ is a point on the surface $S$, i.e. $y \in S$, and
$G_{\mu\nu}(y) = \partial_{\mu} A_{\nu}(y) - \partial_{\nu} A_{\mu}(y)
- ig[A_{\mu}(y), A_{\nu}(y)]$ is the field strength tensor. The
procedure for the derivation of the non--Abelian Stokes theorem in the
form of Eq.(\ref{label8.1}) contains a summation of contributions of
closed paths around infinitesimal areas and these paths are linked to
the reference point $x$ on  the contour $C$  via 
parallel transport operators.  The existence of closed paths linked to
the references point $x$ on the contour $C$ is a {\it necessary} and a
{\it sufficient} condition for the derivation of the non--Abelian
Stokes theorem Eq.(\ref{label8.1}).

Due to the absence of closed paths it is rather clear that the path
integral representation for Wilson loops cannot be applied to the
derivation of the non--Abelian Stokes theorem.  In fact, the
evaluation of the path integral over gauge degrees of freedom demands
the decomposition of the closed contour $C$ into a set of
infinitesimal segments which can be never closed.  Let us prove this
statement by assuming the converse. Suppose that by representing the
path integral over gauge degrees of freedom in the form of the
$n$--dimensional integral (\ref{label2.7}) we have a closed
segment.  Let the segment $C_{x_k x_{k-1}}$ be closed and the point
$x'$ belong to the segment $C_{x_k x_{k-1}}$, $x' \in C_{x_k
x_{k-1}}$. By using Eq.(\ref{label2.2}) we can represent the character
$\chi[U^{\Omega}_r(C_{x_k x_{k-1}})]$ by
\begin{eqnarray}\label{label8.2}
\hspace{-0.2in}&&\chi[U^{\Omega}_r(C_{x_k x_{k-1}})] =
\chi[U^{\Omega}_r(C_{x_i x'})U^{\Omega}_r(C_{x' x_{i-1}})]
=\nonumber\\
\hspace{-0.2in}&&=\chi[\Omega(x_i)U_r(C_{x_i x'})U_r(C_{x'
x_{i-1}})\Omega(x_{i-1})] = \nonumber\\ 
\hspace{-0.2in}&&=d_r\int
D\Omega_r(x')\chi[\Omega_r(x_i)U_r(C_{x_i
x'})\Omega^{\dagger}_r(x')]\,\chi[\Omega_r(x') U_r(C_{x'
x_{i-1}})\Omega(x_{i-1})]=\nonumber\\ 
\hspace{-0.2in}&&=d_r\int
D\Omega_r(x')\chi[U^{\Omega}_r(C_{x_i
x'})]\,\chi[U^{\Omega}_r(C_{x' x_{i-1}})].
\end{eqnarray}
This transforms a $(n-1)$--dimensional integral with one closed
infinitesimal segment into a $n$--dimensional integral without closed
segments. Since finally $n$ tends to infinity there is no closed
segments for the representation for the path integral in the form of a
$(n-1)$--dimensional integral. As this statement is general and valid
for any path integral representation of Wilson loops, so one can
conclude that no further non--Abelian Stokes theorem can be derived
within any path integral approach to Wilson loops.

\section{Conclusion}
\setcounter{equation}{0}

By using well defined properties of group characters we have shown
that the path integral over gauge degrees of freedom of the Wilson
loop which was used in Eq.(2.13) of Ref.[11] for a lattice
evaluation of the average value of Wilson loops can be derived in continuum space--time in 
non--Abelian gauge theories with the gauge group $SU(N)$. The resultant integrand of the path
integral contains a phase factor which is not projected onto Abelian
degrees of freedom of non--Abelian gauge fields and differs substantially from the
representation given in Ref.[3]. The important point
of our representation is the summation over all states of the given
irreducible representation $r$ of $SU(N)$. For example, in $SU(2)$ the
phase factor is summed over all values of the colourmagnetic quantum
number $m_j$ of the irreducible representation $j$ of colour
charges. This contradicts Eq.(23) of Ref.[3], where only term with
the highest value of the colourmagnetic quantum number $m_j = j$ are 
taken into account and the other $2j$ terms are lost. This loss is caused by an artificial regularization procedure
applied in Ref.[3] for the definition of the path integral over gauge
degrees of freedom.

As has been stated by Diakonov and Petrov in Ref.[3] the path integral
over gauge degrees of freedom representing Wilson loops {\it is not
of the Feynman type, therefore, it depends explicitly on how one
``understands'' it, i.e. how it is discretized and regularized}. In
order to {\it understand} the path integral over gauge degrees of
freedom Diakonov and Petrov  [3] suggested a regularization procedure
drawing an analogy between gauge degrees of freedom and dynamical
variables of the axial--symmetric top with moments of inertia $I_{\perp}$
and $I_{\parallel}$.  The final expression for the path integral of the
Wilson loop has been obtained in the limit $I_{\perp}, I_{\parallel}
\to 0$.

In order to make the incorrectness of this expression more transparent
we have evaluated the path integral for specific gauge field
configurations (i) a pure gauge field and (ii) $Z(2)$ center vortices
with spatial azimuthal symmetry.  The direct evaluation of path
integrals representing Wilson loops for these gauge field
configurations has given the value zero for both cases. These results do
not agree with the correct values.

One can show that Eq.(\ref{label5.11}) can be generalized for any contour of a Wilson loop in $SU(2)$
\begin{eqnarray}\label{label9.1}
W_{1/2}(C) &=& \int \prod_{x\in C} D \Omega(x)\,e^{\textstyle ig
\oint_{C}dx_{\mu}\,{\rm tr}[t^3 A^{\Omega}_{\mu}(x)]} = \nonumber\\
&=&\lim_{n\to
\infty}\sum_{j > 0}\Bigg[\frac{j a_j}{2j+1}\Bigg]^n(2j + 1)W_j(C) = 0,
\end{eqnarray}
where $W_j(C)$ in the r.h.s. is defined by Eq.(\ref{label2.1}) in
terms of the path--ordering operator ${\cal P}_C$. Further, the result
(\ref{label9.1}) can be extended to any irreducible representation
of $SU(2)$. Thus, we argue that the path integral
suggested in Ref.[3] to represent Wilson
loops is identically zero for Wilson loops independent on the gauge
field configuration, the shape of the contour $C$ and the irreducible
representation of $SU(2)$.

This statement we have supported by a direct evaluation of the
evolution operator $Z_{\rm Reg}(R_2,R_1)$ defined by Eq.(14) of
Ref.[3], representing the assumption by Diakonov and Petrov for Wilson
loops in terms of the path integral over gauge degrees of freedom. As
we have shown in Sect.\,6 the regularized evolution operator $Z_{\rm
Reg}(R_2,R_1)$, evaluated correctly, is equal to zero. This agrees
with our results obtained in Sects.\,4 and 5. In Sect.\,7 we have
shown that the removal of the oscillating factor from the evolution
operator suggested in Ref.[3] via a shift of energy levels of the
axial--symmetric top is prohibited. Such a shift of energy levels
leads to a change of the starting symmetry of the system from $SU(2)$
to $U(2)$. By virtue of the oscillating factor the Wilson loop
vanishes in the limit $I_{\parallel}, I_{\perp} \to 0$ in agreement
with our results in Sects.\,4, 5 and 6.

We hope that the considerations in Sects.\,4--7 are more than
 enough to persuade even the most distrustful reader that the path
 integral representation for the Wilson derived by Diakonov and Petrov
 by means of {\it special regularization and understanding} of the
 path integral over gauge degrees of freedom is erroneous.

The use of an erroneous path integral representation for Wilson
loops in Ref.[7] has led to the conclusion that for large distances the
average value of Wilson loops shows area--law
falloff for any irreducible representation $r$ of $SU(N)$. 
Unfortunately, this result is not supported by numerical
simulations of lattice QCD [8]. At large distances, colour charges
with non--zero $N$--ality have string tensions of the corresponding
fundamental representation, whereas colour charges with zero
$N$--ality are screened by gluons and cannot form a string. Therefore,
the result obtained in Ref.[7] cannot be considered as {\it
a new check of confinement in lattice calculations} as has been argued
by the authors of Ref.[7].

We would like to accentuate that the problem we have touched in this
paper is not of marginal interest and a path integral, if 
derived by means of an unjustified regularization procedure, would hardly
compute the same physical number as the correct one. We argue that no regularization procedure can lead to
specific dynamical constraints. In fact, the regularization
procedure drawing the analogy with the axial--symmetric top has led to
the result supporting the hypothesis of Maximal Abelian Projection
pointed out by 't Hooft [14]. Any proof of this to full
extent dynamical hypothesis through a regularization procedure and through specific
{\it understanding} of the path integral should have seemed dubious
and suspicious.

Finally, we have shown that within any path integral representation for
Wilson loops in terms of gauge degrees of freedom no non--Abelian
Stokes theorem in addition to Eq.(\ref{label8.1}) can be derived.
Indeed, the Stokes theorem replaces a line integral over a closed
contour by a surface integral with the closed contour as the boundary
of a surface. However, approximating the path integral by an
$n$--dimensional integral at $n \to \infty$ there are no closed paths
linking two adjacent points along Wilson loops.  Thereby, the line
integrals over these open paths cannot be replaced by surface
integrals. Thus, we argue that any non-Abelian Stokes theorem can be
derived only within the definition of Wilson loops through the path
ordering procedure (\ref{label8.1}). Of course, one can represent
the surface--ordering operator ${\cal P}_S$ in Eq.(\ref{label8.1}) in
terms of a path integral over gauge degrees of freedom, but this
should not be a new non--Abelian Stokes theorem in comparison with the
old one given by Eq.(\ref{label8.1}). That is why the claims of
Ref.[3--5] concerning new versions of the non--Abelian Stokes theorems
derived within path integral representations for Wilson loops seem
unjustified.

Discussions with Jan Thomassen are appreciated.

\newpage

\section*{Appendix. Coefficients  $a_j(z)$}

In this Appendix we evaluate the  coefficients $a_j(z)$ of the
expansion (\ref{label4.8}) and show that the completeness condition
given by the series (\ref{label4.16}) converges to unity.

It is convenient to rewrite Eq.(\ref{label4.8}) as follows
$$
e^{\textstyle z\chi_{1/2}[t^3U]}= \sum_{j} a_j(z)\,\chi_j[t^3U]
=a_0(z) + a_{1/2}(z)\,\chi_{1/2}[t^3U]
$$
$$
+
a_1(z)\,\chi_1[t^3U] + a_{3/2}(z)\,\chi_{3/2}[t^3U]  +
a_2(z)\,\chi_2[t^3U] +  a_{5/2}(z)\,\chi_{5/2}[t^3U] + \ldots\eqno(A.1) 
$$
For the evaluation of the coefficients $a_0(z)$, $a_{1/2}(z)$,
$a_1(z)$, $a_{3/2}(z)$, $a_2(z)$ and $a_{5/2}(z)$ we would make use
the parameterization of the matrices $U$ in terms of the Euler angles
$\alpha$, $\beta$ and $\gamma$ ranging over the regions $0 \ge \alpha
\ge 2\pi$, $0 \ge \beta \ge \pi$ and $0 \ge \gamma \ge 2\pi$,
respectively [18]. Then, the Haar measure $DU$
normalized to unity is defined by [18]
$$
DU = \frac{1}{8\pi^2}\,\sin\,\beta\,d\beta\,d\alpha\,d\gamma.\eqno(A.2)
$$
In the notation of Ref.[18] the characters are given by
$$
\chi_{1/2}[t^3U] =- \,i\,\cos\,\frac{\beta}{2}\,\sin\, \frac{\alpha +
\gamma}{2},
$$
$$
\chi_{1}[t^3U] =  -\,
2\,i\,\cos^2\frac{\beta}{2}\,\sin\,(\alpha + \gamma),
$$
$$
\chi_{3/2}[t^3U] =- \,12\,i\,\cos^3\frac{\beta}{2}\,
\sin\,\frac{\alpha + \gamma}{2}\,\cos^2\frac{\alpha + \gamma}{2} +
2\,i\, \cos\,\frac{\beta}{2}\,\sin\frac{\alpha +
\gamma}{2},
$$
$$
\chi_2[t^3U] = -\,
8\,i\,\cos^4\frac{\beta}{2}\, \sin\,(\alpha + \gamma)\,\cos(\alpha +
\gamma)
$$
$$
+ \,i\, (1 + \cos\,\beta)\,(1 - 2\, \cos\,\beta)\sin\,(\alpha +
\gamma),
$$
$$
\chi_{5/2}[t^3U] = -\,
80\,i\,\cos^5\frac{\beta}{2}\,\Bigg( \sin\,\frac{\alpha + \gamma}{2} -
2\,\sin^3\frac{\alpha + \gamma}{2} + \sin^5\frac{\alpha +
\gamma}{2}\Bigg)
$$
$$
- i \,\Bigg(15\,\cos^5\frac{\beta}{2} -
12\,\cos^3\frac{\beta}{2}\Bigg)\,\Bigg( 3\,\sin\,\frac{\alpha +
\gamma}{2} - 4\,\sin^3\frac{\alpha + \gamma}{2}\Bigg)
$$
$$
-\, i\,\Bigg(10\,\cos^5\frac{\beta}{2} - 12\,\cos^3\frac{\beta}{2} +
3\,\cos\,\frac{\beta}{2}\Bigg)\,\sin\frac{\alpha +
\gamma}{2},
$$
$$
\ldots\eqno(A.3)
$$
The coefficients $a_0(z)$, $a_{1/2}(z)$, $a_1(z)$, $a_{3/2}(z)$,
$a_2(z)$ and $a_{5/2}(z)$ are defined by the integrals
(\ref{label4.10}) and (\ref{label4.11})
$$
a_0(z) = \int D U\,e^{\textstyle z\,\chi_{1/2}[t^3U]} =
\frac{1}{8\pi^2}\int\limits^{2\pi}_0d\alpha\int\limits^{2\pi}_0
d\gamma\int\limits^{\pi}_0d\beta\,\sin\,\beta\,e^{\displaystyle -
i\,z\,\cos\,\frac{\beta}{2}\,\sin\,\frac{\alpha +
\gamma}{2}},
$$
$$
a_{1/2}(z) = 4\int D
U\,\chi_{1/2}[t^3U^{\dagger}]\,e^{\textstyle z\,\chi_{1/2}[t^3U]}
=
$$
$$
=\frac{i}{2\pi^2}\int\limits^{2\pi}_0d\alpha\int\limits^{2\pi}_0
d\gamma\int\limits^{\pi}_0d\beta\,\sin\,\beta\,
\Bigg[\cos\,\frac{\beta}{2}\,\sin\, \frac{\alpha + \gamma}{2}
\Bigg]\,e^{\displaystyle -
i\,z\,\cos\,\frac{\beta}{2}\,\sin\,\frac{\alpha +
\gamma}{2}},
$$
$$
a_1(z) = \frac{3}{2} \int D
U\,\chi_{1}[t^3U^{\dagger}]\,e^{\textstyle z\,\chi_{1/2}[t^3U]}
=
$$
$$
=\frac{3i}{8\pi^2}\int\limits^{2\pi}_0d\alpha\int\limits^{2\pi}_0
d\gamma\int\limits^{\pi}_0d\beta\,\sin\,\beta\,
\Bigg[\cos^2\frac{\beta}{2}\,\sin\,(\alpha +
\gamma)\Bigg]\,e^{\displaystyle -
i\,z\,\cos\,\frac{\beta}{2}\,\sin\,\frac{\alpha + \gamma}{2}
},
$$
$$
a_{3/2}(z) = \frac{4}{5}\int D
U\,\chi_{3/2}[t^3U^{\dagger}]\,e^{\textstyle z\,\chi_{1/2}[t^3U]} =
\frac{i}{5\pi^2}\int\limits^{2\pi}_0d\alpha\int\limits^{2\pi}_0
d\gamma\int\limits^{\pi}_0d\beta\,\sin\,\beta\, 
$$
$$
\times\,\Bigg[6\,\cos^3\frac{\beta}{2}\, \sin\,\frac{\alpha +
\gamma}{2}\,\cos^2\frac{\alpha + \gamma}{2} -
\cos\,\frac{\beta}{2}\,\sin\frac{\alpha +
\gamma}{2}\Bigg]\,e^{\displaystyle -
i\,z\,\cos\,\frac{\beta}{2}\,\sin\,\frac{\alpha + \gamma}{2}
},
$$
$$
a_2(z) = \frac{1}{2}\int D
U\,\chi_2[t^3U^{\dagger}]\,e^{\textstyle z\,\chi_{1/2}[t^3U]} =
\frac{i}{16\pi^2}\int\limits^{2\pi}_0d\alpha\int\limits^{2\pi}_0
d\gamma\int\limits^{\pi}_0d\beta\,\sin\,\beta\, 
$$
$$
\times\,\Bigg[8\,\cos^4\frac{\beta}{2}\, \sin\,(\alpha +
\gamma)\,\cos(\alpha + \gamma) - (1 + \cos\,\beta)\,(1 - 2\,
\cos\,\beta)\sin\,(\alpha + \gamma) \Bigg]
$$
$$
\times\,e^{\displaystyle -
i\,z\,\cos\,\frac{\beta}{2}\,\sin\,\frac{\alpha + \gamma}{2}
},
$$
$$
a_{5/2}(z) = \frac{12}{35}\int D
U\,\chi_{3/2}[t^3U^{\dagger}]\,e^{\textstyle z\,\chi_{1/2}[t^3U]} =
\frac{3i}{70\pi^2}\int\limits^{2\pi}_0d\alpha\int\limits^{2\pi}_0
d\gamma\int\limits^{\pi}_0d\beta\,\sin\,\beta\,
$$
$$
\times\,\Bigg[80\,\cos^5\frac{\beta}{2}\,\Bigg( \sin\,\frac{\alpha +
\gamma}{2} - 2\,\sin^3\frac{\alpha + \gamma}{2} + \sin^5\frac{\alpha +
\gamma}{2}\Bigg)
$$
$$
+\Bigg(15\,\cos^5\frac{\beta}{2} -
12\,\cos^3\frac{\beta}{2}\Bigg)\,\Bigg( 3\,\sin\,\frac{\alpha +
\gamma}{2} - 4\,\sin^3\frac{\alpha + \gamma}{2}\Bigg)
$$
$$
 +
\Bigg(10\,\cos^5\frac{\beta}{2} - 12\,\cos^3\frac{\beta}{2} +
3\,\cos\,\frac{\beta}{2}\Bigg)\,\sin\frac{\alpha + \gamma}{2}
\Bigg]\,e^{\displaystyle -
i\,z\,\cos\,\frac{\beta}{2}\,\sin\,\frac{\alpha + \gamma}{2}
},
$$
$$
\ldots \eqno(A.4)
$$
The integration over angle variables gives
$$
a_0(z)= 2\,\frac{J_1(z)}{z}\,,\,a_{1/2}(z)=
8\,\frac{J_2(z)}{z}\,,\,a_1(z) =  0\,,
$$
$$
a_{3/2}(z) = 
\frac{16}{5}\,\frac{J_4(z)}{z},\,\,a_2(z) =  0\,,\,a_{5/2}(z) = 
\frac{72}{35}\,\frac{J_6(z)}{z},\ldots \eqno(A.5)
$$
One can show that for an arbitrary $j$ the coefficient $a_j(z)$ is
defined by
$$
a_j(z) = \left\{\begin{array}{r@{\quad,\quad}l} {\displaystyle
\frac{3(2j+1)}{j(j+1)}\,\frac{J_{2j+1}(z)}{z}} & j =
1/2,3/2,5/2,\ldots,\\\hspace{-0.5in} 0\hspace{0.5in} & j = 1,2,\ldots
\end{array}\right.\eqno(A.6)
$$
This implies the following general formula
$$
\int D U\,\chi_j[t^3U^{\dagger}]\,e^{\textstyle z\,\chi_{1/2}[t^3U]} =
\left\{\begin{array}{r@{\quad,\quad}l} {\displaystyle
(2j+1)\,\frac{J_{2j+1}(z)}{z}} & j =
1/2,3/2,5/2,\ldots,\\\hspace{-0.5in} 0\hspace{0.5in} & j = 1,2,\ldots
\end{array}\right.\eqno(A.7)
$$
The completeness condition (\ref{label4.16}) for arbitrary $z$
takes the form
$$
4\,\frac{J^2_1(z)}{z^2} +
3\sum^{\infty}_{j=1/2}\frac{(2j+1)^2}{j(j+1)}\,
\frac{J^2_{2j+1}(z)}{z^2}=1.\eqno(A.8)
$$
The numerical values of $a_0(z)$, $a_{1/2}(z)$, $a_{3/2}(z)$ and
$a_{5/2}(z)$ for the particular case $z=1$ read
$$
a_0(1)= 0.88\quad,\quad a_{1/2}(1)= 0.92\quad,\quad a_{3/2}(1) =
0.01,\quad,\quad a_{5/2}(1) = 4\times 10^{-5}.\eqno(A.9)
$$
For the completeness condition (\ref{label4.16}) we obtain at $z=1$
$$
a^2_0(1) + \sum^{\infty}_{j =1/2}\frac{1}{3}\,j(j+1)\,a^2_j(1) = a^2_0(1) +
\frac{1}{4}\,a^2_{1/2}(1)+ \frac{5}{4}\,a^2_{3/2}(1)
+\frac{35}{12}\,a^2_{5/2}(1) + \ldots = 
$$
$$
= 0.774 + 0.212
+ 1.25\times 10^{-4} +5\times 10^{-9} + \ldots = 0.986 + \ldots \eqno(A.10)
$$
Thus, this series converges slowly to unity.

\end{document}